\documentclass[aps,prb,singlecolumn,showpacs]{revtex4}
\usepackage{epsfig}
\usepackage{amsfonts,amsmath,amssymb}
\usepackage{graphicx}

\begin{document}

\title{THE CONCEPTUAL HERITAGE OF SUPERCONDUCTIVITY \\
FROM  MEISSNER-OCHSENFELD TO THE HIGGS BOSON}

\author{Julius RANNINGER}

\affiliation{Institut N\'eel, CNRS and Universit\'e Joseph Fourier, \\
 BP 166, 38042 Grenoble cedex 9, France
 \footnote{julius.ranninger@grenoble.cnrs.fr.}}
 
\date{\today}

\begin{abstract}
When first proposed in 1957, the BCS theory for superconductivity, which explained the quasi-totality of its thermodynamic and transport properties, was greeted with great circumspection, before it became the play ground of particle physicists, who largely contributed to understand the deep physics behind this phenomenon. In the course of this undertaking, revolutionizing new concepts in physics were brought to light, such as (i) the "physical significance  of the phase" of a quantum mechanical wave function whose role in force-transmitting gauge fields is to control the interaction between elementary particles in a current conserving manner (ii) "spontaneous symmetry breaking"  and  its related to it collective Nambu-Goldstone modes, which encode the basic symmetry properties of a quantum vacuum and the associated to them conserved quatities, (iii) The "Anderson-Higgs mechanism" and its associated to it, but so far experimentally unconfirmed "Higgs field", which provides the introduction of massive force transmitting gauge fields and ultimately the mass of elementary particles. These concepts were vital in consolidating the standard model for elementary particles and presented the final answer to what  distinguishes a superconducting from a 
non-superconducting state:  "electromagnetic gauge symmetry breaking", whatever the microscospic mechanism for that might be. We illustrate here how these concepts gradually emerged, once the basic features, which characterize superconductivity - the Meissner-Ochsenfeld magnetic field screening and the fundamental London equations explaining them in a phenomenological way - were established. 

\vspace{2cm}
keywords: Superconductivity, electromagentic gauge invariance breaking, Proca equation, Nambu-Goldstone modes, massive photons, Anderson-Higgs mechanism
\end{abstract} 

\maketitle

\newpage
\section{Introduction}
Celebrating anniversaries of important scientific discoveries are occasions to 
reflect on the fundamental issues which were discussed at the time and which  
have since been largely forgotten, if not put deliberately aside in order to present 
popularized pedagogical descriptions in an easily accessible fashion to a broad 
scientific community. By doing this, the subtlety and  
profoundness of the arguments, which initially led to the comprehension of 
such discoveries are  frequently lost. To remind oneself of the original 
intellectual effort necessary  to grasp fundamentally new  and basically 
unexpected physical
phenomena, is important and not just for historical reasons. Commemorating 
the discovery of superconductivity by Kamerlingh Onnes \cite{Kamerlingh-Onnes-1911} 
in April 1911, one faces the problem of presenting a subject 
about which everything, which possibly could have been said, had been said.  Yet, 
this enigmatic phenomenon of superconductivity  has lost nothing of its original, 
still intriguing and to some extent poorly understood fundamental 
features. \cite{Anderson-1986,Leggett-Sols-1991,Leggett-2011}
Our desperate struggle, for now a quarter of a century, to unravel 
the mysteries of the  so-called high temperature cuprate superconductors, 
\cite{Bednorz-Mueller-1986} witnesses that. I shall review here, the seemingly 
"never ending" hurdles which had and still have to be overcome in  order to grasp 
the phenomenon of superconductivity and demonstrate how our illustrious 
predecessors in this 
process  acquired a deep physical understanding of quantum mechanics and 
the subsequently emerging quantum field theory, which was instrumental for the
achievements in  particle physics throughout the second half of last century. 

The present essay is based on a talk, which I presented in April 2011 to a 
broad scientific 
community, composed of solid state, nuclear and particle physicists, in a centennial 
celebration of the discovery of superconductivity, organized by the French Physical 
Society in Grenoble. The discussions, comments and questions, which this presentation 
sparked off have largely helped me to clarify many still nebulous points. I hope that 
with this review on superconductivity traversing the twenty's century, newcomers 
into the field might look at it as a challenging  subject which still has many surprises 
up its sleeves. 

The prime task of physics has always been and always will be, to reduce the 
 diversity of physical phenomena, which we are confronted with by our 
experimental observations, to  a strict minimum of basic laws of nature which 
might govern it. They 
represent conservation laws which control the dynamical processes of particles in 
their interaction with each other as well as in their decay processes. Such 
laws  are formulated in terms of conserved quantities, which reflect fundamental 
symmetry properties of our universe. In classical Newtonian mechanics, 
space-time symmetries reflect the conservation of linear and angular momentum and 
energy. In classical electrodynamics they assure the conservation of charge 
through the invariance of the Maxwell's equations under a  space-time 
correlated Lorentz transformation and a concomitant gauge transformation 
of the electromagnetic field tensor. Trying to comprehend our physical 
world along such lines, Galileo 
Galilei conjectured that the dynamics of our sub-lunar and  supra-lunar world should  
be controlled by one and the same mechanism.  Isaac Newton reconciled 
terrestrial motion with that of celestial one on a quite general level.  Michael 
Faraday, discovering electromagnetic induction, paved the way to  Maxwell's 
unification of electricity and magnetism. A turning point came to this approach
when Einstein inversed this so-far applied principle by hypothesizing  rather than 
by deriving general symmetry properties from experimental observations. From 
his proposition of a global and continuous space-time symmetry, he concluded that 
the laws of our physical world were controlled  by  special relativity, rather 
than classical mechanics. 

Attempting to understand the physics on a subatomic level and to predict and 
classify the "elementary" particles: the Leptons (the electrons, the muons 
and the tau mesons, together with their corresponding neutrinos) and the six 
quarks, interacting with each other via force-transmitting gauge bosons (the 
massless photons and gluons and the massive Z$^0$ and W$^{\pm}$ bosons), 
quantum field theory introduced a set of new conservation laws. They 
complemented the classical laws, related to space-time symmetries in form of the 
CPT theorem \cite{Lueders-1954} and introduced the socalled "internal symmetries". 
The CPT theorem, involving charge conjugation, parity 
and time reversal transformations, requires that any combination of all three 
leaves any elementary particle as well as interaction invariant. The 
internal symmetries are at the origin of conservation laws  which 
control the different quantum numbers, associated to the elementary particles and their 
force-transmitting gauge fields: The conservation of Barion and Lepton number, 
electric charge, hypercharge, Isospin, quark-colour and quark-flavour.

By incorporating those "internal symmetries" in a covariant Lagrangian formulation 
and subsequently searching for solutions which 
break these symmetries, enabled particle physicists 
to predict elementary particles together with  their force transmitting
excitations, which could be rendered visible in nuclear decay processes. The 
radioactive $\beta$ decay represented such an occasion for that: it breaks
parity symmetry but not PC and conserves the Lepton and Barion 
number. \cite{Lee-Yang-1956} Sheldon Glashow, Steven Weinberg and 
Abdus Salam, \cite{Glashow-1961,Weinberg-1967,Salam-1968}  subsequently 
showed that, as a consequence the electromagnetic forces (controlling the  
interaction between electrons) and the weak nuclear forces (involved in this
parity breaking process), although being widely different in strength and range, 
represent just different manifestations of a common underlying electro-weak 
force. During the evolution of the universe, such different manifestations of 
a basically unique physics to start with, emerge through a hierarchy of spontaneous 
symmetry breaking processes, the outcome  of which characterizes the state of our 
present day world. 

The understanding of what makes a metal to become a superconductor and not merely 
a perfect resistance-less conductor, was crucial in arriving at the level 
of understanding such basic principles and laws of nature. Fundamentally new
concepts in physics were unearthed in this process. They acquired their
profound physical implications and full understanding really only  once Bardeen, 
Cooper and Schrieffer in their BCS theory \cite{BCS-1957}  had understood and 
described the quasi totality of the basic thermal and  transport properties
of superconductors. Their microscopic approach focused on the superconducting
gap in the single-particle spectrum of the electrons, considering the gap to  play
the role, at that time, of an order parameter. \cite{Bardeen-1956} The real 
robust features of the superconducting state, which substantially revised our 
concepts in physics at large, however became apparent from its macroscopic manifestations: 
the Meissner-Ochsenfeld effect \cite{Meissner-Ochsenfeld-1933}, 
the flux quantization \cite{Deaver-Fairbank-1961,Doll-Naebauer-1961}  and the 
Josephson effect, \cite{Josephson-1962} in conjunction with their description  
in terms of the phenomenological Ginzburg-Landau \cite{Ginzburg-Landau-1950} 
theory. This theory derived its input from the Meissner-Ochsenfeld effect and
Fritz London's early conjecture \cite{FLondon-1937} of a truly macroscopic wave 
function describing the current carrying superconducting state, tantamount to 
a macroscopic diamagnetic molecule. The year after that, when superfluidity of 
$^4$He was discovered \cite{Wilhelm-1935,Allen-Misener-1938,Kapitza-1938} and 
the connection between superconductivity and superfluidity became apparent, 
Fritz London's picture for superconductivity, was determinant for pinning down the 
robust features behind this phenomenon: The first one was its connection to a 
Bose-Einstein condensation, \cite{Bose-1924,Einstein-1925} which for the first time
acquired a true physical significance. The second and even more important 
feature, as far as superconductivity is concerned, was its intrinsic 
perfect diamagnetism. It is related to persistent super-currents, driven by a 
self-sustaining mechanism which relies on a qualitative modification of 
the classical Maxwell equations of electromagnetism. As Fritz and Heinz London  
showed, these equations had to be complemented by what has been termed the 
London equations \cite{FHLondon-1935}. The implications of that  however go far 
beyond the explanation of the Meisser-Ochsenfeld screening of a magn{eq:supercurrent}etic field 
from the interior of a superconductor. It led to fundamentally 
new concepts, which were determinant for our understanding of 
quantum mechanics and its evolution into quantum field theory and particle physics, such as: 

i)  {\bf The phase of matter waves - gauge bosons and gauge fields}.

In the early days of quantum mechanics, in the beginning of the last century, 
the phase of de Broglie matter waves was considered at best as a "redundant degree 
of freedom" - an indeterminable gauge of the wave function of no particular
physical significance. Yet, a deep physical meaning could not be denied for a 
similarly indeterminable gauge of the four-component electromagnetic 
vector-scalar field of the Maxwell equations. The invariance 
of the Maxwell equations under gauge transformations are rooted
in an underlying conserved quantity: the electric charge of the particles. 
When charge carriers interact with such an electromagnetic field, the 
Lagrangian describing this interaction has to be gauge invariant. The albeit 
arbitrary phase of the quantum mechanical wave functions of the charged particles
then becomes  irrevocably tied to the gauge of the vector-scalar components
of the electromagnetic field in which they are embedded. 
The electromagnetic properties of superconductors, permitted us to visualize 
the physical effect of gauge transformations, relating the dynamics of massive 
charge carriers (the electrons) to that of massless gauge bosons, playing the role
of the local force-transmitting excitations which accompany them.

It is generally attributed to  Wolfgang Pauli, \cite{Pauli-1940} for  having 
introduced the  local gauge theory approach into particle physics. The local gauge 
in quantum field theories  - the Yang-Mills gauge 
theories \cite{Yang-Mills-1954} - assures the conservation laws controlling 
the dynamics of elementary particles together with their 
force transmitting gauge bosons. The basic idea behind that was the same as that 
which Emmy Noether had laid down before in her "Noether's 
theorem". \cite{Noether-1918,Travel-1971} (For a modern discussion of it, see 
ref. \onlinecite{Byers-1999}) It states that the conservation of
quantities, such as momentum, angular momentum and energy arises from basic 
symmetries of our classical physical world: homogeneity in space and time and 
isotropy. From that alone follows that the dynamics in such systems must be
invariant with respect to continuous global transformations of the space-time 
and rotational coordinates.

The "quantum field theory" extended these so-called Poincar\'e symmetries
of classical space-time translational invariance to "internal symmetries", 
characterizing  elementary particles and their 
associated gauge bosons. The latter present the quantized excitations of the 
force transmitting fields, similar to those of photons in the electromagnetic
field which accompany the electrons. Gauge field theories, 
culminating in the standard model, assume as an outset a highly symmetric 
quantum vacuum, composed of a multitude of different degrees of freedom. Following
the BIG BANG, the universe evolved by expanding and thereby cooled down. In
this course of events, a hierarchy of symmetry breaking processes occurred 
which ultimately led to our presently observed state of forces, widely differing 
in strength and range:  the electromagnetic, the weak and the
strong nuclear forces. The gravitational force remains outside of such a 
scheme. Integrating it into it would require to  treat general relativity 
(which considers the  mass as a localized in space object) and quantum mechanics 
(which considers it as a spatially extended wave packet), on the same 
footing. Presently, it is not known how this could be achieved.
 
ii) {\bf Spontaneous symmetry breaking - Nambu-Goldstone modes.} 

Spontaneous symmetry breaking (SSB) implies that a physical system with an internal 
global continuous symmetry, resulting in a certain conserved quantity, 
has a ground state in which this symmetry is broken without any external 
perturbation causing it. SSB was conjectured by Phil Anderson \cite{Anderson-1958bis}
in connection with the  breaking of the continuous rotational symmetry 
of the phase in the superconducting ground state. The infinite multiplicity of 
its degenerate ground state leads to the emergence of long wave length massless 
spin-0 scalar boson collective excitations, known as Nambu-Goldstone modes (under 
the condition that the interactions in such a system are short ranged), 
which reestablish the original global symmetry in a dynamical 
fashion.\cite{Nambu-1960,Nambu-2008} Nambu showed that these modes 
contribute to the charge current in a way, without which it would 
not be conserved. We shall illustrate in Sec. VI how this comes about, on the
basis of a concrete example, that of a charged superfluid Bose liquid. 

The Nambu-Golstone modes distinguish themselves from massless excitations 
in systems with unbroken or weakly broken  symmetries in the sense, that they
encode the structure of the originally symmetry-unbroken situation. An example 
for that is the fluctuating microwave background, which emerged after the BIG BANG 
in form of a SSB state of the "inflationary" fully symmetric behavior of the our 
universe as it existed before it.  

It is generally accepted to credit Nambu as the most instrumental physicist for 
having transposed the concepts of the SSB in superconductivity to Lorentz 
covariant gauge field theories, as recalled by one of his collaborators, 
Jona Lasinio \cite{Jona-Lasinio-2003} as well as by Steven
Weinberg \cite{Weinberg-1986,Weinberg-2008}. Its outcome was the rigorous  formulation of the 
"Goldstone theorem", which stated that upon breaking an exact global 
continuous symmetry, necessarily leads to the emergence of massless Nambu-Goldstone  bosons \cite{Goldstone-1961,Goldstone-1962}. 
This theorem played a considerable role in the construction of the standard model for 
elementary particles, whose major aim  was to bring order and classification 
into the flood of experimental discoveries of potentially elementary
particles and to reduce the fundamental laws of physics acting between them
to a strict minimum. Its first success was the unification of the electromagnetic
and the weak nuclear forces by Sheldon Glashow \cite{Glashow-1961}, Steven 
Weinberg \cite{Weinberg-1967} and Abdus Salam \cite{Salam-1968}, to which we 
have alluded above. But this necessitated  having to overcome yet another 
and major obstacle: The force-transmitting gauge fields involved in the $\beta$
decay, which was at the basis of this unification claim, experimentally turned
out to have mass. This was in flagrant dis-accord with the concept of the 
intrinsic masslessness of gauge bosons. It implied the violation of gauge 
invariance and hence called into question the current conservation.  Once more, 
the phenomenon of superconductivity, played a decisive role in resolving this riddle, through the "Anderson-Higgs mechanism".

iii) {\bf The emergence  of massive gauge bosons - The Anderson-Higgs mechanism.} 

The experiment, which  pinned down once and for all the singular behavior 
of superconductors, was the screening of an externally applied magnetic 
field from the interior of a persistent current-carrying superconducting hollow Pb 
sphere, observed by Walther Meissner and Robert Ochsenfeld in 
1933. \cite{Meissner-Ochsenfeld-1933} More then 20 years had passed since the 
discovery of superconductivity, when this observation finally led the way to 
its ultimate understanding. It required a qualitative modification of 
the electromagnetic field which accompanies such a persistent current as soon as
the superconducting state sets in below a critical temperature $T_c$. 
This current is self-sustained, persistent and geometrically restricted to a 
thin layer near the surface of this sphere, which prevents any electromagnetic 
field to penetrate in its interior deeper than the Meissner-Ochsenfeld
penetration  depth.  In 1935 Fritz and Heinz London showed that in order to 
describe such a situation, the local electromagnetic field, which accompanies 
this super-current,  materializes if the current-density source terms 
of the Maxwell equations are self-determined. They have to be related back 
to those electromagnetic fields via two phenomenological equations, the socalled 
London equations. \cite{FHLondon-1935} Effectively, this amounts to transform 
the Maxwell equations into Proca equations \cite{Proca-1936}, which describe
electric and magnetic fields which have a mass, inversely proportional to 
the penetration depth.

On a microscopic level this requires turning massless transverse photons of 
the electromagnetic field in vacuum, i.e., outside the superconductor, into 
massive excitations when they enter the superconducting material and interact 
with the circulating current. The conservation of this current is encoded in the characteristic properties of the Nambu-Goldstone mode of the 
spontaneously broken symmetry of the condensed state of the superconductor. 
But coupling this current to an electromagnetic field, locally breaks the gauge 
invariance of that latter. In order to assure the conservation of this current,
a self-sustaining feedback effect between the current carried by the charge
carriers and that carried by the accompanying electromagnetic vector field, 
establishes itself. This goes at the expense of the freely propagating features 
of the photons which accompany this current locally and which in this process 
are turned into massive excitations. This process has been termed the Anderson-Higgs 
mechanism.\cite{Anderson-1963} Its generalization to relativistic gauge 
theories was done by Higgs \cite{Higgs-1964} and independently by Englert and 
Brout \cite{Englert-Brout-1964} and  Guralnik, Hagen and
Kibble. \cite{Guralnik-Hagen-Kibble-1966} It made it possible to circumvent
the intrinsic masslessness of the  gauge fields when they are associated via a 
local gauge transformation to a massless Nambu-Goldstone mode. This makes this 
mode disappear as a physical particle, but resurrects it in form of a helicity-0 
spin-1 vector boson with mass - presenting in a more complex scenario than presented here,  the actively searched after Higgs Boson. 

These concepts evolved gradually as the understanding of superconductivity 
progressed. The reason, why it has taken such a long time for that is rooted in 
our traditional reductionist approach to physical problems \cite{Anderson-1972}. 
The superconducting state is caused by electromagnetic gauge symmetry 
breaking, independent on any specific microscopic scenario. A concept which 
was difficult to  accept. One should remember that up to and 
even still slightly after the publication of the BSC theory in 1957, it was tabu to 
consider that the gauge (the phase) of a matter wave could possibly have any 
physical role to play in superconductivity. The proposition that its 
"redundant" character could be lost
by acquiring a definite value, which leads to gauge symmetry breaking, took 
even longer to be accepted. To show how this break-through of our thinking 
evolved throughout the last century, let me  highlight below the main milestones 
in this process, illustrating the thorny step by step advances in understanding 
the phenomenon of superconductivity, which has taken half a century after 
its discovery.

 In order to make the meaning of the superconducting state clear, I shall in 
 Sec. II illustrate its differences with respect to an ideal conductor. The 
 main characteristics of a superconductor - the superconducting gap and the 
 Meissner-Ochsenfeld magnetic field screening - will be discussed in Sec III.
 Combining the
 Maxwell equations with the phenomenological London equations in Sec IV, we
 shall highlight its effect on the introduction of a longitudinal mode in 
 the electromagnetic excitations. They are forbidden in vacuum because of
 the Lorentz condition, but which becomes redundant for massive electromagnetic 
 field equations. The absorption of the Nambu-Goldstone mode in a SSB state of
 the superconductor and its reappearance in form of a massive helicity-0 spin-1
 vector boson field will be discussed in terms of the Ginzburg-Landau 
 macroscopic formulation of superconductivity in Sec V. Finally, in Sec VI we
 shall treat this feature on the  basis of a microscopic generic model  
 for superconductivity in terms of a charged Bose gas in a manifestly gauge
 invariant manner and derive generalized London equations for 
 spatial and temporarily varying electromagnetic source fields, which leads 
 us up to the Proca equations for the electromagnetic fields accompanying
 and sustaining the super-current. In the final section VII we conjecture, how the  
 fundamental concepts which originated from our understanding the BCS 
 superconductivity might have to be complemented and integrated into a wider 
 scheme, when addressing the physics of  the high $T_c$ cuprate superconductors.
 We advance some  preliminary ideas based on the concept of "emerging phenomena"
 whose dynamical stability are assured  by such features as "quantum protection".

\section{The perfect conductor and the myth of the frozen-in magnetic field, 1911-1933}
In 1908 Kamerlingh Onnes succeeded in  liquefying gaseous $^4$He below 4.22 K. This  
earned him the 1913 Nobel prize and gave him a worldwide monopoly to explore 
physical phenomena at such low temperatures. It was in this way that superconductivity 
was discovered - without that anybody would have foreseen it - in 1911. In 
this experiment a magnetic field was applied traversing a ring
of frozen Mercury. The temperature then was reduced to below some critical 
temperature $T_c$, 
in this case 4.15 K. When subsequently the external field was shut off, a 
persistent resistance-less current (super-current) manifested itself in the ring.
It was evidenced in form of  a magnetic flux threading through the ring, which 
was equal in strength to that of the field initially applied to it. It led to the
long standing official doctrine that superconductivity was induced by the 
externally applied field having gotten frozen into the sample before cooling it 
down to below $T_c$. This interpretation was supported by considering 
superconductors as  resistance-less "perfect conductors", as illustrated by 
Schoenberg, \cite{Schoenberg-1952} following an earlier study by Becker, Heller 
and Sauter. \cite{Becker-1933}  The argument goes as follows.

In a "perfect conductor",  a super-current ${\bf J}_s = n_s q_s \dot{\bf x}$, with 
$\dot{\bf x}$ representing its velocity, is accelerated in an applied local
electric field ${\bf E}$ according to:
\begin{equation}
\label{eq:supercurrent}
\partial_t {\bf J}_s = {n_s q_s^2 \over m_s} {\bf E},
\end{equation}
where $n_s$, $m_s$ and $q_s$ are the  effective density, mass and charge of 
the  particles making up such a super-current.
Using Ampere's and Faraday's laws: $\nabla \times {\bf B} = {\bf J}_s$ 
and $\nabla \times {\bf E} = - {1 \over c} \partial_t {\bf B}$, with ${\bf B}$ and ${\bf E}$ denoting 
the local magnetic and electric fields,  we have  
\begin{equation}
{1 \over c}\partial_t \nabla \times {\bf B} =  \lambda^{-2}_L {\bf E} \Longrightarrow 
\nabla^2 {\bf E} = \lambda^{-2}_L {\bf E},
\label{eq:perfectconductor}
\end{equation}
with $\lambda_L = \sqrt{m_s c/ n_s q_s^2} = \sqrt{1/n_s r_s}$. $r_s = q_s^2/m_s c$ plays the role of the radius of the charge carriers with charge
and mass given  by  $q_s$ and $m_s$. 
 
Requiring this super-current to be homogeneous ($\nabla {\bf E} \equiv 0$), it 
follows from Eq. \ref{eq:perfectconductor} that it should flow without any 
voltage difference, i.e., ${\bf E} \equiv 0$  everywhere in the sample. 
Yet, considering that such a super-current is triggered by applying a magnetic 
field, things become bothering.

Taking the curl of Eq. \ref{eq:supercurrent} and using Faraday's  law we have:
\begin{equation}
\partial_t \left[ \nabla \times {\bf J}_s + {n_s q_s^2 \over m_s c} {\bf B}\right] = 0.
\label{eq:Londoneq_2bis}
\end{equation}
Eliminating subsequently ${\bf J}_s$  via Ampere's law and taking into account 
that $\nabla {\bf B} \equiv 0$ it follows that:
\begin{equation}
\partial_t \left[\nabla^2 - {1 \over \lambda^2_L} \right] {\bf B} \equiv 0.
\label{eq:Meissnerscreening-bis}
\end{equation}
Solving this equation for a half-infinite system with its surface orthogonal to the 
z-direction and for an externally applied field ${\bf B}_0 = [B^x_0, 0, 0]$, 
polarized parallel to this surface, one has
\begin{equation}
\partial_t B^x(z.t) = \partial_t B^x_0(t) exp(-z/\lambda_L).
\label{eq:Fluxrelaxation}
\end{equation}
This implies that any temporal variation of the externally applied field, 
like switching it on or off, will result in a relaxation of this perturbation 
to the system, which will have no effect on the flux in the sample, deeper than
a distance $\lambda_L$, as measured from the surface. In other words, such
a perturbation has no effect on any pre-existing magnetic flux inside the 
sample - a situation, which has unphysical consequences.

When a magnetic field is switched on in a perfect conducting ring in the 
resistanceless state below a certain temperature $T_c$, it "kicks off" 
a circulating super-current, following  Lenz's law and prevents in this 
way the initially
applied magnetic field to penetrate into the bulk of the sample. When subsequently 
the external field is shut off, the circulating current stops 
and with it, its accompanying magnetic flux in the sample. The system returns to 
its initial state with no current flowing.

The situation is qualitatively different, when applying the magnetic field  above 
$T_c$ in the resistive  state of a "perfect conductor". The magnetic
flux then traverses the bulk of the sample. Upon lowering the temperature 
to below $T_c$,  the flux line  distribution in the bulk remains unchanged 
and no super-current 
is being set off. In order to generate a super-current to flow, one 
must switch off the externally applied field.  Following the relation,
Eq. \ref{eq:Fluxrelaxation}, which controls the relaxation of such a magnetic 
flux, its distribution inside the material will tend to that which it was,  when 
the system was first exposed to a magnetic field above $T_c$. Hence the conclusion  
of a frozen-in magnetic field driving the super-current.

The resistanceless state thus depends on the
order  in which this state is approached: either by first lowering the temperature
and then switching on the field or the inverse. It means that the superconducting 
state would not be a truly thermodynamically stable state. This highly 
unsatisfactory conclusion fortunately turned out to be wrong. But it has taken 
more than twenty years after the discovery of superconductivity that this was 
finally recognized, thanks to the experimental findings in 1933  by Meissner and 
Ochsenfeld, \cite{Meissner-Ochsenfeld-1933} which stipulated that: (i) The field 
inside a superconductor is identically zero. The super-current is self-sustained by 
the field which it creates via  Ampere's law and screens any 
externally applied field. (ii)  The order, in which the temperature and the
magnetic field is varied in inducing the super-current carrying state and its 
associated characteristic screening features, is irrelevant. 

This finding by Meissner and Ochsenfeld established without 
ambiguity the true thermodynamically stable state nature for superconductivity 
and its second order phase transition into a non-superconducting metallic
phase. A superconductor hence proved to be more than simply a "perfect conductor".
The observation of the "frozen-in magnetic field" in those early years could have 
been due a variety of different unfortunate circumstances: the shape of the 
sample, magnetic supercooling and flux trapping as the magnetic field is switched off.

\section{The two fundamental properties of superconductors: 
The Macroscopic Quantum Phase Coherence and the Single Particle Gap} 
By 1928 the advances in Quantum Mechanics had sufficiently progressed to provide 
one with a theoretical description of de Broglie matter waves for the electrons
in Arnold Sommerfeld's and Felix Bloch's  theory  of 
metals. \cite{Sommerfeld-1928,Bloch-1928} By 1933, 
Ehrenfest \cite{Ehrenfest-1933} provided us with the concept of second order 
phase transitions, which evolved into   Lev Landau's concept of an order 
parameter. \cite{Landau-1937}   Curiously, one of
the most relevant concepts for describing  the superconducting state, i.e.,  
its coherent macroscopic quantum state, proposed by Erwin Schroedinger in 
1926, \cite{Schroedinger-1926} remained totally unnoticed in the solid state 
community. Schroedinger's proposition captured  the classical features of a 
quantum system, such as to satisfy optimally the correspondence 
principle with a minimal Heisenberg quantum uncertainty. The concept of 
macroscopic quantum coherence resurfaced only after the observation of the total loss 
of viscosity in the flow of liquid $^4$He through pores and in general any obstruction, 
below a certain temperature $T_{\lambda}$ = 2.18 K. \cite{Wilhelm-1935,Allen-Misener-1938} The flow  being controlled exclusively by its inertia, led
Kapitza \cite{Kapitza-1938} to call this feature: superfluidity. The observed 
specific heat cusp at the transition into this state gave it the name 
"lambda transition". It strongly resembled that expected for the specific heat of a 
free Bose gas, undergoing a condensation of the majority of bosonic $^4$He
atoms into a single quantum state. Its transition temperature 
$T_{BEC} = (2 \pi \hbar^2/ 1.8 m_{He} k_B) n^{2/3}$, with $m_{He}$ given by the mass
of the $^4$He atoms and n $\simeq 10^{22}/cm^3$ being their density, comes 
surprisingly close 
to the observed value of $T_{\lambda}$. The frictionless flow of superfluid 
$^4$He in a tube without any difference in pressure on the two ends, is the 
analog of the resistanceless conductivity,  i.e., the persistent flow of an 
electric current in a superconductor in the limit where the driving electric 
field goes to zero. Driving an electric super-current by 
applying a magnetic field in the Meissner-Ochsenfeld experiment found  its 
analog much later in  the rotating bucket experiment \cite{Hess-Fairbank-1967} in $^4$He.

Fritz London \cite{FLondon-1938} and independently 
Tisza \cite{Tisza-1938}  recognized that superfluid Helium and 
superconducting metals must contain basically the same underlying physics, 
inherent to Bose quantum liquids. Their ground state therefore
should be describable by a single macroscopic matter wave function.  It implied 
generalizing  Schroedinger's phonon coherent state, accounting  for the 
macroscopic features of a quantum harmonic oscillator, to that of an equivalent coherent 
state of massive bosonic $^4$He atoms.  This
concept ultimately led to the highly successful phenomenological Ginzburg-Landau 
theory of superconductivity \cite{Ginzburg-Landau-1950}. Penrose and 
Onsager \cite{Penrose-Onsager-1956} consolidated this concept by attributing to
such a macroscopic wave wavefunction the intrinsic property of "long range 
off-diagonal order", which, according to C. N. Yang, \cite{CNYang-1962} encodes the 
concept of "spontaneous gauge symmetry breaking".

The steady state of a persistent homogeneous super-current, circulating in a ring, 
which follows from such a macroscopic wave function for a condensate,  suggested 
to Fritz London \cite{FLondon-1937} that any particle making up the super-current 
must be correlated to any other one, independent on the spatial difference between 
two. This pointed to electron pairing and effective charge carriers having charge 
2e. Fritz London moreover conjectured that such a current should  behave similar to
that of a rotating  macroscopic diamagnetic molecule under the influence of an 
applied magnetic field. Since the quantum 
mechanical spatial correlations between any two particles in the current on a 
ring could be considered as  independent on their relative distance, it implied that
the conjugate momentum of it had to be fixed within the limits of  Heisenberg's 
uncertainty principle. Fritz London conjectured from that, that the rigidity 
of the macroscopic wave function of a super-current on a ring derived from
pairing correlations in reciprocal rather than real space, limited to the momenta 
of the particles which could participate in such a process, i.e., those close to 
the Fermi surface since all others are blocked by the  Pauli exclusion principle. 
This insight was instrumental for John Bardeen and his collaborators to 
construct ultimately their macroscopic coherent quantum state capable of
describing dynamical pairing of electrons rather than bound pairs, which would mimic 
the superfluidity of bosonic $^4$He atoms. That latter scenario  was 
followed by Schafroth and his collaborators Blatt and Buttler, \cite{Schafroth} 
who tried to relate superconductivity to a Bose Einstein condensation of 
diamagnetic bound electron pairs. What
eventually decided in favor of John Bardeen's approach,
was the experimentally observed gap in the electronic single-particle spectrum. 
Its difference with respect to an insulating gap is encoded in the specific
structure of the macroscopic quantum state for the BCS wave function with (i) 
its coherence factors describing the intrinsically Fermionic system and (ii) 
its single particle excitations, derived by  Bogoliubov \cite{Bogoliubov-1958} and 
independently Valatin, \cite{Valatin-1958} which are quantum superpositions 
of negatively charged electrons and positively charged holes.

Keesom's and Kok's early measurement of the specific heat \cite{Keesom-Kok-1932}
of a superconducting tin sample showed a jump in the specific heat at $T_c$, 
which  indicated the opening of a gap in the electron spectrum as one entered
the superconducting  phase. A  definite proof for it came later in experiments 
on the change-over in the thermal conductivity \cite{Hulm-1950} upon entering
the superconducting state with decreasing temperature. Since a super-current does 
not transport heat, this experiment permitted to follow the variation
with temperature of this gap in the electron spectrum. It led John
Bardeen \cite{Bardeen-1955} to conjecture that the gap 
should play the role of an order parameter. Immediately after the publication of 
the microscopic BCS theory, \cite{BCS-1957} which aimed to account for  
the thermodynamic and transport properties of superconductors, this conjecture was 
elaborated by Gorkov. \cite{Gorkov-1959} He succeeded to make the connection with the 
phenomenological Ginzburg-Landau theory \cite{Ginzburg-Landau-1950} and its complex  
order parameter, given by the macroscopic condensate wave function, whose 
modulus represents the value of the gap, but whose phase controls the  onset 
of superconductivity. 

\section{The Meissner-Ochsenfeld effect and the London Equations and how they 
influenced the electromagnetic properties}

In interpreting the Meissner-Ochsenfeld experiment, Fritz and Heinz 
London \cite{FHLondon-1935} stipulated that: 
   
\vspace{0.5cm}

I) The friction-less conduction of super-carriers with some effective charge 
$q_s$, mass $m_s$ and  charge/mass density $n_s$ manifests itself  in 
the acceleration of a charge-current which, in the limit  ${\bf E} \rightarrow 0$ 
is described by  the "first London equation", which is identical to Eq. \ref{eq:supercurrent}, for perfect conduction without Meissner-Ochsenfeld screening:                                                                                                                                                                                                                                                                                                                                                                                                                                                                                                                                                                                                                                                                                                                                                                                                                                                                                                                                                                                                                                                                                                                                                                                                                                                                                                                                                                                                                                                                                                                                                                                                                                                                                                                                                                                                                                                                                                                                                                                                                                                                                                                                                                                                                                                                                                                                                                                                                                                                                                                                                                                                                                                                                                                                                                                                                                                                                                                                                                                                                                                                                                                                                                                                                                                                                                                                                                                                                                                                                                                                                                                                                                                                                                                                                                                                                                                                                                                                                                                                                                                                                                                                                                                                                                                                                                                                                                                                                                                                                                                                                                                                                                                                                                                                                                                                                                                                                                                                                                                                                                                                                                                                                                                                                                                                                                                                                                                                                                                                                                                                                                                                                                                                                                                                                                                                                                                                                                                                                                                                                                                                                                                                                                                                                                                                                                                                                                                                                                                                                                                                                                                                                                                                                                                                                                                                                                                                                                                                                                                                                                                                                                                                                                                                                                                                                                                                                                                                                                                                                                                                                                                                                                                                                                                                                                                                                                                                                                                                                                                                                                                                                                                                                                                                                                                                                                                                                                                                                                                                                                                                                                                                                                                                                                                                                                                                                                                                                                                                                                                                                                                                                                                                                                                                                                                                                                                                                                                                                                                                                                                                                                                                                                                                                                                                                                                                                                                                                                                                                                                                                                                                                                                                                                                                                                                                                                                                                                                                                                                                                                                                                                                                                                                                                                                                                                                                                                                                                                                                                                                                                                                                                                                                                                                                                                                                                                                                                                                                                                                    \begin{equation}
{\partial \over \partial t} {\bf J}_s = {n_s q_s^2 \over m_s} {\bf E}.
\label{eq:Londoneq_1}
\end{equation}

II) The exclusively transverse component of such a steady state super-current 
${\bf J}_s$, i.e., $\nabla \times {\bf J}_s \neq 0$, prevents any magnetic flux 
from entering the bulk of a superconducting material beyond a certain penetration 
depth $\lambda_L$ according to the "second London equation",

\begin{eqnarray}
\nabla \times {\bf J}_s = - {n_s q_s^2 \over m_s c} {\bf B}.
\label{eq:Londoneq_2}
\end{eqnarray}

The effect of these two premises, although not spelled out at that time, is that 
the steady state of the super-current and its associated to it local electromagnetic 
field, as we understand it now, sustain each other mutually in the sense that: (i) 
When a magnetic field is applied to the material above the critical temperature $T_c$,  
a super-current is spontaneously put into circulation as soon as the temperature
is reduced to below $T_c$. 
Shutting off this external field, the super-current disappears. (ii) A material 
which has already been cooled to below  $T_c$, exhibits a super-current as 
soon as an external magnetic field is applied to it. The super-current, accompanied
by its electromagnetic screening field, persists until the external field  is shut off.

\vspace{0.5cm}

As we have seen in  Sec. II, a superconducting state, triggered by an external magnetic 
field below $T_c$, can adequately be accounted for its frictionless conduction 
in terms of a "perfect conductor", described by the "first London equation", 
Eq. \ref{eq:Londoneq_1}. It however can 
neither explain the superconducting state in a field-cooled experiment nor 
the screening of the magnetic flux from the interior of the material. In order 
to remedy this situation, Fritz and Heinz London \cite{FHLondon-1935} conjectured 
in 1935 that Eq. \ref{eq:Londoneq_2bis} for a "perfect conductor" should be replaced by 
the more stringent 
relation, expressed in their "second London equation", Eq. \ref{eq:Londoneq_2}.  
Eq. \ref{eq:Meissnerscreening-bis}, describing the "perfect conductor" then gets replaced by
\begin{equation}
\left[\nabla^2 -{1 \over \lambda^2_L}\right]{\bf B} = 0
\label{eq:Meissnerscreening}
\end{equation}
in the case of a superconductor. If one requires the magnetic flux to be 
homogeneous in the interior of the superconductor, it follows from 
Eq. \ref{eq:Meissnerscreening} that  ${\bf B}$ must be identically zero in the 
bulk of the material beyond a certain penetration depth $\lambda_L$. Fritz and
Heinz London conjectured from this that charged particles in the superconducting 
and in the normal state interact differently with an electromagnetic field. As 
we shall see below, this is related to the breaking of the local electromagnetic 
gauge invariance in the superconducting state.

Felix Bloch suggested that the "second London equation" results from the fact that 
the charged particles (making up the  super-current) in the presence of a 
magnetic field ${\bf B({\bf x})} = \nabla \times {\bf A}({\bf x})$ should be 
described by the canonical momenta 
${\bf p} = m_s \dot{\bf x} + (q_s/c){\bf A}({\bf x})$, rather than by the usual 
simple kinetic contribution $m_s\dot {\bf x}$, where ${\bf A}({\bf x})$ denotes 
the local vector potential. In the ground state, the total canonical momentum 
of the ensemble of charge carriers has to be equal to zero: 
$\langle 0| {\bf p} |0 \rangle = 0$
\begin{equation} 
\langle 0|{\bf J}_s({\bf x}) | 0 \rangle = {n_s q_s^2 \over m_s c}{\bf A}({\bf x}),
\label{eq:Londoneq_2ter}
\end{equation}
from which follows the "second London equation". Eq. \ref{eq:Londoneq_2ter} 
visibly violates electromagnetic gauge invariance 
(${\bf A}({\bf x}) \rightarrow {\bf A}({\bf x}) +\nabla \Lambda({\bf x})$) and 
that therefore charged particles in the superconducting state, contrary to its 
normal state, defy the 
conservation of the charge-current. It took several years after the 
publication of the BCS paper in 1957 that this problem was remedied. 
As Nambu \cite{Nambu-1960} showed, the kinetic part of the charge 
current has to be complemented by a contribution 
${\bf J}_{coll} = \nabla f({\bf x},t)$, which arises from collective
modes, described by $f({\bf x},t)$ and satisfying the wave equation 
\begin{equation} 
(\nabla^2 -{1 \over \alpha^2} \partial_t^2)f({\bf x},t) = -2 \Delta \Psi^{\dagger} \tau_2 \Psi({\bf x},t).
\label{eq:Jcoll-Nambu}
\end{equation}
$\Delta$ denotes the gap value, $\Psi({\bf x},t)$ the electron-hole spinor wave 
function and $\alpha$ the plasma frequency of the free electron gas.
This puts back Felix Bloch's Ansatz, Eq. \ref{eq:Londoneq_2ter}  into a 
gauge invariant form. How this happens in detail will be addressed in sections 
V and VI below.

For the moment let us pursue the  classical phenomenological description
of superconductivity, based on the London equations and illustrate how the
electromagnetic field is modified, when it enters the superconductor and couples 
to its circulating currents. In a superconducting metal,  the total current is
composed of two contributions: ${\bf J} = {\bf J}_s + {\bf J}_n$, where 
${\bf J}_n = \sigma {\bf E}$ denotes that part of it which flows with a 
finite resistivity $1 / \sigma$ and ${\bf J}_s$ represents the super-current, 
which above $T_c$ is identically zero. 

Complementing Maxwell equation in vacuum
\begin{eqnarray}
\nabla \times\, {\bf B} = {\bf J} + {1 \over c} \partial_t {\bf E},  \quad 
\nabla \times  \; {\bf E} = -{1 \over c} \partial_t {\bf B}, \quad
\nabla {\bf B} = 0, \quad \nabla {\bf E} = \rho, 
\label{eq:MaxwellEqs}
\end{eqnarray}
whose corresponding wave equations for the electric and magnetic fields are  given by
\begin{eqnarray}
\left(- \nabla^2 + ({1 \over c})^2 \partial^2_t \right) {{\bf E} \brack {\bf  B}} = 
{4 \pi \over c} {-{1 \over c} \partial_t {\bf J} -  \nabla \rho \brack  \nabla \times {\bf J}}
\label{eq:MaxwellEqs-bis}
\end{eqnarray}
with the London equations, Eqs. \ref{eq:Londoneq_1} 
and \ref{eq:Londoneq_2}, we obtain wave equations 
\begin{eqnarray}
\left(-\,\nabla^2\, + {1 \over c^2}\partial^2_t\, + 
{m^2_0 \, c^2 \over \hbar^2} + {1 \sigma \over c} 
\partial_t \right) \{{\bf B}, {\bf E}, {\bf J}\} = 0, 
\label{eq:ProcaEqus}
\end{eqnarray}
not just for the usual transverse excitations of the electric and magnetic 
fields ${\bf E}$ and ${\bf B}$ but also for the longitudinal excitations 
associated to the current ${\bf J}$. In order to derive this result we have
put  the charge source terms in the Maxwell equations Eqs. \ref{eq:MaxwellEqs} 
equal to zero. In the Meissner-Ochsenfeld 
experimental set up, the sole external source initiating the circulation of a 
current comes from a magnetic field, i.e., from the transverse component of the 
vector potential and not from any distribution of localized 
charge carriers which would produce such an electric field source term. 

All of these modes are massive, with a 
rest-mass $m_0$ given by $m^2_0c^2/\hbar^2 = n_s q^2_s/m_s c$.
These equations present a generalization of the Maxwell equations to finite 
mass photons and were initially conjectured by Proca \cite{Proca-1936} on general
field theoretical grounds in view of generalizing the Klein-Gordon equation for 
massive spin-0 scalar Boson fields to fields describing massive spin-1 
vector Boson fields. 

Let us now highlight the difference of the electromagnetic properties between the normal non-superconducting materials, described by the standard Maxwell wave equations  Eqs. \ref{eq:MaxwellEqs-bis} and the superconducting materials, described by the Proca equations, Eqs \ref{eq:ProcaEqus}. 

We begin with the non-superconducting case for which the standard Maxwell equations Eqs. \ref{eq:MaxwellEqs} apply and rephrase these equations in their  relativistically covariant form
\begin{eqnarray}
\nabla^2 \phi({\bf x},t) - {1 \over c^2} \partial_t^2  \phi({\bf x},t)  &=&  
- {1 \over c}\partial_t \left( \nabla  {\bf A}({\bf x},t)  + 
{1 \over c} \partial_t \phi({\bf x},t)\right ) - \rho({\bf x},t) \nonumber \\
\nabla^2  {\bf A}({\bf x},t) - {1 \over c^2} \partial_t^2  {\bf A}({\bf x},t)
&=& \nabla\left( \nabla  {\bf A}({\bf x},t)  + 
{1 \over c} \partial_t \phi({\bf x},t)\right ) - {\bf J}({\bf x},t),
\label{eq:MaxwellwaveEqs}
\end{eqnarray}
after having cast the physical electric and magnetic fields into a gauge 
invariant form by introducing the scalar and vector fields 
[$\phi({\bf x},t),{\bf A}({\bf x},t)$] with
${\bf B}({\bf x},t) = \nabla \times {\bf A}({\bf x},t)$ and 
${\bf E}({\bf x},t) = - \nabla \phi({\bf x},t) - 
\frac{1}{c}\partial_t {\bf A}({\bf x},t)$. 
Eqs. \ref{eq:MaxwellwaveEqs} present historically the first example of a unification 
of different forces: the electric and magnetic forces. Their invariance under the electromagnetic gauge transformations
$[\phi({\bf x},t),{\bf A}({\bf x},t)] \rightarrow [\phi'({\bf x},t) = 
\phi({\bf x},t)- \frac{1}{c}\partial_t \Lambda({\bf x},t),{\bf A}'({\bf x},t) =  
{\bf A}({\bf x},t) + \nabla \Lambda({\bf x},t)]$  ($\Lambda({\bf x},t)$ being 
an arbitrary function of ${\bf x}$ and $t$) results from an underlying symmetry 
of electromagnetism, which is associated to the conservation of charge.
In order to demonstrate that, let us operate with $\frac{1}{c}\partial_t$ 
onto the first of these  two equations, Eqs. \ref{eq:MaxwellwaveEqs}, and with
$\nabla$ onto the second one. When subsequently summing the two equations, we obtain 
the continuity equation 
$\frac{1}{c}\partial_t \rho({\bf x},t) +\nabla {\bf J}({\bf x},t) = 0$, which 
asserts that the Maxwell equations are charge-current conserving and that the 
symmetry associated to this conservation law is that of electromagnetic 
gauge invariance. Since the Maxwell equations are  also invariant under the 
space-time Lorentz transformations, $[x' = x, y' = y, z' = (z-vt)/\sqrt{1 - (v/c)^2}, t' = (t -(vz/c^2)) / \sqrt{1 - (v/c)^2}$, it requires (see for details 
Ref. \onlinecite{Panofsky-Phillips-1955}) that the charge-current conservation,
i.e., the continuity equation, must be invariant under Lorentz transformations. 
This leads to the  socalled "Lorentz condition"
\begin{eqnarray}
 \nabla  {\bf A}({\bf x},t)  + {1 \over c} \partial_t \phi({\bf x},t) = 0,
 \label{eq:Lorentzcond}
\end{eqnarray}
which correlates $\phi({\bf x},t)$ and ${\bf A}({\bf x},t)$ and renders the Maxwell equations fully symmetric 
\begin{eqnarray}
\nabla^2 \phi({\bf x},t) +{1 \over c^2}\partial^2_t\phi({\bf x},t) &=&
- \rho({\bf x},t) \nonumber \\
\nabla^2 {\bf A}({\bf x},t) - {1 \over c^2} \partial_t^2  {\bf A}({\bf x},t) &=& 
- {\bf J}({\bf x},t).
\label{eq:MaxwellwaveEqs-bis}
\end{eqnarray}
It expresses the fact that the charge density and the current density together 
with their associated scalar and vector fields are merely different manifestations
of a single and unique underlying physics.

The wave character of Eqs. \ref{eq:MaxwellwaveEqs-bis} is contained in their 
homogeneous solutions, which can be parametrized by $\phi({\bf x},t)$ = 
e$_{\phi} e^{- i({\bf q}\cdot{\bf x} - \varepsilon t)}$ and ${\bf A}({\bf x},t) = 
{\bf e}_{\bf A} e^{- i({\bf q}\cdot{\bf x} - \varepsilon t)}$. e$_{\phi}$ 
and ${\bf e}_{\bf A}$ present the components of the polarization vectors of the scalar, respectively 
the vector field and ${\bf q}$ and $\varepsilon$ are the frequency and the wave 
vector of these modes. Substituting this Ansatz into the Lorentz condition, 
Eq. \ref{eq:Lorentzcond}, we find 
$\varepsilon$e$_{\phi} + {\bf q}\cdot {\bf e}_{\bf A} = 0$. 
This leaves us with three independent modes out of the four we have started with. But 
the Lorentz condition itself, having to be invariant under the electromagnetic 
gauge transformations, permits us to reduce the number of independent modes 
even further. We can choose for that any one of the components of 
[$\phi({\bf x},t),{\bf A}({\bf x},t)]$ to be eliminated. Let us choose it 
such that the time component $\phi({\bf x},t)$ vanishes. This is achieved
by transforming Eq. \ref{eq:Lorentzcond} by the gauge transformation
$\phi({\bf x},t) \rightarrow \phi'({\bf x},t) = \phi({\bf x},t) - 
\frac{1}{c} \partial_t \Lambda({\bf x},t) =0$,  with $\Lambda({\bf x},t)$ =
i(e$_{\phi}/\varepsilon) e^{- i({\bf q} \cdot {\bf x} - \varepsilon t)}$. Together with 
the constraint $\varepsilon$e$_{\phi} + {\bf q} \cdot {\bf e}_{\bf A} = 0$, we 
are thus left with ${\bf q} \cdot {\bf e}_{\bf A} = 0$, which  leaves us with 
two independent  modes, orthogonally polarized  to the direction
of the wave propagation ${\bf q}$. This  of course only reiterates the general quantum
field theoretical argument that the excitations of a spin-1 vector gauge boson field,  
such as the electromagnetic field, can neither have finite mass nor a 
longitudinal component. The electromagnetic force fields, thus having infinite
range, implies that transverse modes of the electromagnetic field propagate freely in 
vacuum. The  magnetic field, transported by these modes therefore freely 
transverses a metal in its non-superconducting state, apart from a minimal
Landau diamagnetic response to it.

This is no longer the case when this metal becomes a superconductor. Bringing 
the electromagnetic field into contact with a  supercurrent, the  transverse 
photons can no longer propagate freely, as evidenced by the Meissner-Ochsenfeld magnetic field  screening. As we have seen in the previous section the Maxwell  equations then have to be replaced by the Proca equations, Eqs. \ref{eq:ProcaEqus}. Rephrasing these equations in  terms of the vector and scalar fields, in analogy to the Maxwell equations in vacuum, Eqs. \ref{eq:MaxwellwaveEqs} and putting the source terms $\propto \rho ,{\bf J}$ equal to zero, we obtain the quantum field theory equations  for spin-1 vector gauge fields:
 \begin{eqnarray}
\left( \nabla^2  +  {m_0^2 c^2 \over \hbar^2}\right) \phi({\bf x},t)
- {1 \over c^2} \partial_t^2  \phi({\bf x},t) + 
{1 \over c}\partial_t \left( \nabla  {\bf A}({\bf x},t)  + {1 \over c} 
\partial_t \phi({\bf x},t)  \right) &=& 0 \nonumber \\
\left( \nabla^2  + {m_0^2 c^2 \over \hbar^2}\right){\bf A}({\bf x},t) - {1 \over c^2} \partial_t^2  {\bf A}({\bf x},t)- \nabla\left( \nabla  {\bf A}({\bf x},t)  + 
{1 \over c} \partial_t \phi({\bf x},t)\right ) &=& 0.
\label{eq:MaxwellEq-massive}
\end{eqnarray}

The introduction of the mass term in these equations qualitatively changes the 
spectral properties of the excitations of these gauge fields. In order to see that, let us investigate what happens to the polarization vectors of 
$\phi({\bf x},t)$ and ${\bf A}({\bf x},t)$ when passing from the massless 
spin-1 vector boson fields to massive ones. Operating onto the first of these
two equations, Eqs. \ref{eq:MaxwellEq-massive}, with $-\frac{1}{c}\partial_t$ and 
on the second equation with $\nabla$ and subsequently summing the two, we have 
 \begin{eqnarray}
 {m_0^2 c^2 \over \hbar^2}\left( {1\over c}\partial_t \phi({\bf x},t)+
 \nabla {\bf A}({\bf x},t)\right)  = 0.
 \label{eq:Lorentzcondition-bis}
 \end{eqnarray}
This implies that the Lorentz condition is automatically satisfied and therefore 
can not serve to reduce the number of independent modes from four to three, as was 
the case of the Maxwell equations, Eqs. \ref{eq:MaxwellwaveEqs} describing 
massless photons. Implementing Eq. \ref{eq:Lorentzcondition-bis},  
Eqs. \ref {eq:MaxwellEq-massive} are in fact intrinsically equivalent, i.e.,
 \begin{eqnarray}
 \left( \nabla^2 - {1 \over c^2} \partial_t^2 + {m_0^2 c^2 \over \hbar^2}\right) \phi({\bf x},t)
 &=& 0 \nonumber \\ 
\left( \nabla^2 -  {1 \over c^2} \partial_t^2 + {m_0^2 c^2 \over \hbar^2}\right){\bf A}({\bf x},t)  
 &=& 0.
\label{eq:MaxwellEq-massive-bis}
\end{eqnarray}

But using, as before, the Ansatz  $\phi({\bf x},t)$ = 
e$_{\phi}e^{- i({\bf q}\cdot{\bf x} - \varepsilon t)}$ and ${\bf A}({\bf x},t) = 
{\bf e}_{\bf A} e^{- i({\bf q}\cdot{\bf x} - \varepsilon t)}$, we obtain from 
Eq. \ref{eq:Lorentzcondition-bis}: \;
$[\varepsilon$ e$_{\phi} - {\bf q} \cdot {\bf e}_{\bf A}  = 0]$,
which reduces the number of modes from four to three independent modes. But
unlike in the case of the Maxwell equations in vacuum, 
Eqs. \ref{eq:MaxwellwaveEqs}, which are invariant under gauge transformations, 
Eqs. \ref{eq:MaxwellEq-massive-bis} are not any longer. Hence 
no gauge transformation can reduce the number of independent modes any further.
We thus have two massive transverse  circularly polarized modes with 
polarization vectors ${\bf e}_{\perp} = \frac{1}{\sqrt 2}[0, 1, \pm i, 0]$ and 
one longitudinal massive mode with a polarization vector
${\bf e}_{\parallel}= (c^2 q_z^2 + \varepsilon^2)^{-1/2}[c q_z, 0, 0, \varepsilon]$, which can be considered as a helicity-0 component  spin-1 vector gauge boson. 

The equations  \ref{eq:MaxwellEq-massive-bis}, which result from combining the
Maxwell equations with the London equations, represent  the deep physical 
content of the Meissner-Ochsenfeld effect. Transposing these intricate 
electromagnetic features of the  spontaneously broken symmetry of the phase 
locked state of a superconductor, onto that of the matter field, described by
a polarized quantum vacuum in contact with a massless spin-1 boson vector 
field, culminated in the emergence of massive gauge  fields in the standard 
model of elementary particle physics via the Anderson-Higgs mechanism.

Before illustrating in the next two sections how mass production in the 
force-transmitting excitations of the electromagnetic field comes about, let
us look at a simpler situation, namely that described by the relativistic 
wave equation for spin-0 particles, i.e.,the Klein-Gordon wave equation
\begin{equation}
 \left( \nabla^2 - {1 \over c^2} \partial_t^2 + {m_0^2 c^2 \over \hbar^2}\right) \Phi({\bf x},t) = 0.
\end{equation}
This equation was proposed around 1926 simultaneously by different authors as a relativistic generalization of the Schroedinger equation for a single particle. The mass production here can be envisaged as arising from a breaking of a continuous translational symmetry, such as that experienced in a chain of atoms 
coupled with each other on nearest neighbor sites.  This is a perfectly continuous translational invariant system, whose modes are the acoustic phonons with an energy given by $\omega_q = v q$. But when such a chain of atoms is embedded in a substrate, in such a way that the displacement of the atoms are coupled to  fixed position in a substrate, the translational symmetry is broken and the excitations will have acquired a finite mass. The Hamiltonian for such a scenario is given by
\begin{equation}
H_{Lattice} = \sum_{<i,j>}{1\over 2}\left[m \dot u^2_i+\omega^2 (u_i - u_j)^2  + \omega^2_o u^2_i\right].
\label{eq:HlatticeDiscrete}
\end{equation}
In the continuous limit it becomes
\begin{equation}
H_{Lattice} = \int dx {1 \over 2}\left[m (\dot \Phi(x,t))^2 + v^2 (\nabla \phi)^2(x,t) + \omega^2_o \Phi^2(x,t)\right].
\label{eq:HlatticeContinuum}
\end{equation}
$v=\omega a$ and $\Phi(x,t) = u_i(t)\sqrt a$. The interatomic 
distance $a$ of the lattice has to be taken equal to zero as we approach the 
continuum limit.  The excitations of this system are then described by 
the dynamical lattice deformations, having the energies 
$\omega_q = \sqrt{ v^2 q^2 + \omega_o^2}$, with $\omega^2_o$ playing the role 
of the mass which is induced by the polarization of the acoustic phonons due to their 
coupling to the broken symmetry of the substrate. As we shall see in the next two 
sections, the mass of the photons accompanying the 
super-current feel the rotational symmetry breaking of the superconducting phase, 
which, like in the example given here, couples to the translational massless 
modes of the electromagnetic field and thereby renders them massive.

\section{The emergence of massive gauge Bosons - the Anderson-Higgs mechanism and its  field theoretical treatment}

By the mid fifties, the Lorentz covariant gauge-field theoretical approach 
in elementary particle physics was well on the way. It was based on the concept
of a quantum vacuum, composed of a multitude of degrees of freedom, 
which, via the fluctuations of symmetry broken polarized states, could give these 
degrees of freedom a true physical significance. The resulting excitations represent 
the fermionic elementary particles: six quarks and the six leptons: 
the electron, the muon and the tau, together with their corresponding neutrinos. 
A particular role in this approach is played by the dynamical breaking of continuous  
symmetries, via the so called spontaneous symmetry breaking 
(SSB). \cite{Nambu-1960,Nambu-2008} It results in the emergence of bosonic collective 
Nambu-Goldstone modes. The associated to them excitations are manifest in form 
of conserved currents, which  reflect the basic conservation laws in particle physics to
which we have alluded to in the Introduction and which derive from "internal symmetries" of the 
quantum vacuum. The Goldstone theorem stipulates that the Nambu-Golstone modes should be 
inherently massless. \cite{Goldstone-1961,Goldstone-1962}.
Inspired by the concept of Cooper pairs, made out of itinerant  electrons in the
BCS theory, led Nambu and Jona-Lasinio \cite{Nambu-JonaLasinio-1961a,
Nambu-JonaLasinio-1961b} to propose Dirac fermions as composites of mesons and 
nucleons as a result of chiral symmetry breaking.

The introduction  of quantized force-transmitting gauge fields (the photon for 
the electromagnetic forces, the gluons for the strong nuclear forces and 
the $W^{\pm}$ and $Z^0$ for the electro-weak forces), which accompany the
elementary particles, assure that the conservation laws are respected in 
scattering and decay processes. By construction, these gauge fields are  
massless bosons.  But with the exception of the massless photons, the remaining 
gauge bosons all are massive. In the example of the $\beta$ decay, the  mass of
the force-transmitting $W^{\pm}$ and $Z_0$ gauge bosons amount to roughly
$80 \; GeV$. The intrinsic masslessness of the force-transmitting gauge boson, 
let alone that of the fermionic elementary particles in such an approach, 
presented a considerable stumbling block for these gauge field theories. 

A way out of this dilemma was proposed by Julian Schwinger. \cite{Schwinger-1962} He 
suggested that local current conservation of 
a conserved quantity could still be maintained provided one associates to it a
gauge transformation, which simultaneously acts onto the gauge fields,   
accompanying such currents. It is in this way that gauge fields acquire  mass.

Phil Anderson \cite{Anderson-1963} came up with a physical realization for 
such a scenario in terms of a non-relativistic analog:  a plasma of a free
electron gas, in which gauge invariance as well as particle conservation are 
assured, even though the associated to it gauge photons are not 
collective low frequency modes. The
propagating transverse and longitudinal plasma modes, with frequencies above the
plasma frequency, play in this scenario the role of Schwinger's conjectured massive 
vector gauge bosons. The density fluctuations of such an electron plasma display 
an infinite polarizibility which is tantamount to the polarizibility of the 
matter field, i.e., the quantum vacuum fluctuations in particle physics. 
The longitudinal plasma mode, 
corresponding to the helicity zero gauge boson in Lorentz gauge invariant field 
theory, would be absent if one could switch off its gauge coupling to the 
density fluctuations of the electron gas, respectively the symmetry broken 
vacuum fluctuations in quantum field theory. But since the longitudinal plasma 
modes couples via gauge transformations to a background electromagnetic gauge field, it  thereby switches on the otherwise forbidden longitudinal component of this gauge field. 
It is this, which subsequently induces a mass in the transverse modes and renders the 
passage of transverse electromagnetic waves in such a medium  opaque, when
their frequency is below that of the plasma mode. This scenario
for introducing massive excitations into the  Yang-Mills gauge theories
was subsequently transposed onto Lorentz covariant gauge field theories 
by Higgs \cite{Higgs-1964} and independently by Englert and
Brout \cite{Englert-Brout-1964} as well as  Guralnik, Hagen and 
Kibble, \cite{Guralnik-Hagen-Kibble-1966} which significantly helped to 
consolidate the standard model.

The Anderson-Higgs mechanism is based on a  self-consistent dielectric mechanism 
in metals in which current fluctuations ${\bf J}({\bf q},\varepsilon)$, monitored 
by the the plasma oscillations, act as source terms  in the associated electromagnetic field ${\bf A}^m({\bf q},\varepsilon)$
\begin{equation} 
A^m({\bf q},\varepsilon) = {1 \over \left[c^2 q^2 - \varepsilon^2 \right]} 
J^m({\bf q}\,\varepsilon).
\end{equation}
But since the current fluctuations themselves are monitored by the linear response 
$K^{mn}({\bf q},\varepsilon)$ to this electromagnetic field
${\bf A}^m({\bf q},\varepsilon)$, 
it follows from the classical dielectric response in metals \cite{Nozieres-Pines-1958} that
\begin{equation} 
J^m({\bf q},\varepsilon) = -K^{mn}({\bf q},\varepsilon) A^n({\bf q},\varepsilon),
\end{equation}
where $\varepsilon$ and ${\bf q}$ denote the frequency and wavevectors of these fields.
In the static limit ($\varepsilon \rightarrow 0$), gauge invariance imposes 
the structure of the kernel:  
$K^{mn}({\bf q},0) = (q^m q^n - \delta^{mn}q^2) K(q^2)$.
With the choice of the London gauge, i.e., 
${\bf q} \cdot {\bf A}({\bf q},\varepsilon) = 0$,  
this reduces to $K^{mn}({\bf q},0)=-\delta^{mn}n_e e^2 / m_e c$. $n_e$ and $m_e$ 
denote the density and mass of 
the charge of the electrons. Substituting this expression 
for $K(q^2)$ into the above two coupled equations, results in the wave equations  
for the three components of the electromagnetic vector field

\begin{equation}
\left[c^2 q^2 - \varepsilon^2 + {n_e e^2 \over m_e c}\right]
A^m({\bf q},\varepsilon) = 0,
\end{equation}
when in contact with the electrons which make up the plasma. 

The  Anderson Higgs mechanism, in the example of an electron plasma, hinges on two 
conditions: (i) The  fluctuating electron plasma (the matter-field, mimicking 
the quantum vacuum with a corresponding  Nambu-Goldstone mode in quantum 
field theory), must have an intrinsic infinite polarizibility before being coupled 
to any force-transmitting gauge field. (ii) The physical manifestations of the
in principle unobservable underlying symmetry of the  matter field are apparent
in the conserved currents, which correlate the dynamics of the matter-field and the 
gauge fields, \cite{Anderson-1958bis} in an overall electromagnetic gauge 
invariant manner. 

Let us conclude this section by the field theoretical formulation of the 
Anderson-Higgs mechanism for converting massless gauge fields into massive ones.
We follow for that purpose the well presented discussion of it, given in 
ref. \onlinecite{Negele-Orland-1998}.  Our illustration here is based on a 
paradigm for superconductivity in terms of a complex scalar field, which 
plays the role of an order parameter of a SSB state and which is locally 
coupled to an electromagnetic field. This paradigm represents the 
Ginzburg-Landau phenomenology of a superconductor, described by a macroscopic 
condensate wave function
\begin{eqnarray} 
\Psi^*(x) = e^{-i\Theta(x)}|\Psi(x)| \quad \Psi(x) = e^{+i\Theta(x)}|\Psi(x)|.
\end{eqnarray}
We assume the SSB superconducting state to be  parametrized by a Mexican hat potential 
\begin{eqnarray}
V[\Psi^*(x),\Psi(x)] = -|\alpha|\Psi^*(x)\Psi(x) + {|\beta| \over 2}\left[\Psi^*(x)\Psi(x)\right]^2
\end{eqnarray} 
and that the condensate wave function is coupled to a locally gauge invariant
four component electromagnetic gauge field $A^{\mu}(x)$.
The density fluctuations of such a superfluid condensate  $\delta\Psi(x)$ around a
fixed average amplitude $|\Psi(x)| = \Psi_0 =\sqrt{n_s}=\sqrt{\alpha/\beta}$ and  
the phase fluctuations $\delta\Theta(x)$ around a symmetry broken state, given 
by an arbitrary but fixed phase,  $\Theta_0$, are parametrized by 
\begin{eqnarray} 
\Psi(x)&=& e^{i (\Theta_0 + \delta\Theta(x))}(\Psi^0 + \delta\Psi(x)) \nonumber \\
\Psi^*(x) &=& e^{-i (\Theta_0 + \delta\Theta(x))}(\Psi^0 + \delta\Psi(x)). 
\end{eqnarray}
The Lagrangian for the superconductor matter-field, coupled to the electromagnetic 
gauge field in a relativistically covariant Minkowski metric 
[$x^{\mu} = \{ct, x, y, z\}, \; x_{\mu} = \{ct, -x, -y, -z\}$, is (in natural units $\hbar = c = 1$) given by 
\begin{eqnarray}
L [\Psi^*(x),\Psi(x), A(x)] &=& \left[ (\partial^{\mu}+ ieA^{\mu}(x))\Psi(x)
\right]^*\left[(\partial_{\mu} + ieA_{\mu}(x))\Psi(x)\right] + V[\Psi^*(x),\Psi(x)]
- {1 \over 4} F^{\mu\nu} F_{\mu\nu}.
\label{eq:Lagrangian_1}
\end{eqnarray}
$L[A(x)] = - {1 \over 4} F^{\mu\nu} F_{\mu\nu}$ denotes the Lagrangian  for 
the electromagnetic field in terms of the electromagnetic field tensor 
$F^{\mu \nu} = \partial^{\mu} A^{\nu}(x) - \partial^{\nu}A^{\mu}(x)$, where 
$A^{\nu}(x)$  satisfies the Maxwell equations: 
\begin{eqnarray}
{\delta L[A(x)] \over \delta A^{\mu}(x)} = 0 \, \rightarrow \, \partial^{\nu} F_{\nu\mu} = 0,
\end{eqnarray}
and whose solutions are given by two orthogonally polarized transverse modes. 
The longitudinal mode 
is prohibited by gauge invariance of these equations under  the transformations
$A_{\mu}(x) \rightarrow A'_{\mu}(x) = A_{\mu}(x) + \partial_{\mu} \Lambda(x)$ 
and imposing the Lorentz condition $\partial_{\mu} A^{\mu}(x) = 0$, as we have illustrated in the previous section IV.

In the absence of any electromagnetic gauge field (setting 
${\bf A}(x) = F^{\mu \nu}(x) = 0$ in the above Lagrangian), we obtain the dynamics 
of the phase and amplitude fluctuations by developing this Lagangian to second 
order in  $\delta\Psi(x)$ and $\delta\Theta(x)$ around the stationary solution 
$|\Psi(x)|^2 = \Psi_0^2 = n_s = -\alpha/\beta$ and $\Theta(x) = \Theta_0$ denoting 
an arbitrary but fixed phase between $0$ and $2\pi$:
\begin{eqnarray}
\partial^{\mu}\partial_{\mu} \delta\Theta(x) &=& 0 \\
\left(\partial^{\mu}\partial_{\mu} - 2 \alpha \right) \delta\Psi(x)= 0 \; &\Rightarrow& 
\left( - E^2 + p^2 +m_0^2\right)\delta\Psi(x) = 0.
\end{eqnarray}
Its solutions  are given by a massive amplitude mode $\delta\Psi(x)$ with a
mass $m_0^2 = 2 |\alpha| = 2 n_s \beta$ and a massless Nambu-Goldstone
mode $\delta\Theta(x)$.

This massless Nambu-Goldstone mode is eliminated, when the spontaneously 
broken matter field is coupled to the electromagnetic field through the 
first term in Lagrangian, Eq. \ref{eq:Lagrangian_1}. A simple gauge 
transformation
\begin{eqnarray}
\Psi(x) &\Rightarrow& \widetilde {\Psi}(x) = e^{-i \Phi(x)}\Psi(x) \Rightarrow \Psi_0 + \delta\Psi(x) \nonumber
\\
\Psi^*(x) &\Rightarrow&  \widetilde {\Psi}^*(x) =   e^{+ i \Phi(x)}\Psi^*(x) \Rightarrow \Psi_0 + \delta\Psi(x) \nonumber\\
A^{\mu}(x) \Rightarrow \widetilde {A}^{\mu}(x) &=& A^{\mu}(x) + (1/e)\partial^{\mu} \Phi(x) \Rightarrow A^{\mu}(x) + (1/e)\partial^{\mu} \Theta(x)
\end{eqnarray}
can do that, given the fact that the electromagnetic field tensor is locally 
gauge invariant. The original Lagrangian, Eq. \ref{eq:Lagrangian_1}, then is 
transformed into
\begin{eqnarray}
L [\widetilde{\Psi}^*(x),\widetilde{\Psi}(x), \widetilde{A}(x)] &=& 
\left[ \partial^{\mu} -  ie\widetilde{A}^{\mu}(x)\right] \left(\Psi_0 + \delta\Psi(x)\right)
 \left[\partial_{\mu} + ie\widetilde{A}_{\mu}(x)\right] \left(\Psi_0 + \delta\Psi(x)\right) \\
&& -\alpha \left(\Psi_0 + \delta\Psi(x)\right)^2 - 
{\beta \over 2}\left(\Psi_0 + \delta\Psi(x)\right)^4 - {1 \over 4} \widetilde{F}^{\mu\nu} \widetilde{F}_{\mu\nu}.
\label{eq:Lagrangian_2}
\end{eqnarray} 
The massless phase fluctuations of the Nambu-Goldstone mode,  characterizing 
the spontaneously  broken symmetry of the superconductor matter-field before
its coupling to the electromagnetic fields, have thus dropped out in
this procedure. They have  been absorbed - "eaten up" - by the excitations 
of a renormalized electromagnetic field $\widetilde{A}_{\mu}(x)$.
Variation of  $L [\widetilde{\Psi}^*(x),\widetilde{\Psi}(x), \widetilde{A}(x)]$ 
with respect to $\{ \delta \Psi(x), \widetilde{A}(x)\}$ leads to 
 \begin{eqnarray}
\left(\partial^{\mu}\partial_{\mu} - m_0^2\right)\delta\Psi(x) &=& 0 \\
{m_0^2 e^2 \over \beta} \widetilde{A}^{\mu}(x) + \partial_{\nu} F^{\nu\mu}(x) \Rightarrow;
\left(\partial_{\nu}\partial^{\nu} - {m_0^2 e^2 \over \beta}\right)
\widetilde{A}^{\mu}(x) &=& 0,
\end{eqnarray}
with the second equation representing the Proca equation for massive electromagnetic 
fields. It describes the emergence of the  massive longitudinal mode and the 
transformation of the massless transverse photons into massive ones. For 
obtaining this equation, we  have exploited the identity 
\begin{eqnarray}
 \partial _{\nu}F^{\nu\mu}(x) = \partial_{\nu}\partial^{\nu} A^{\mu}(x) - \partial^{\mu}\partial_{\nu}A^{\nu}(x) \equiv \partial_{\nu}\partial^{\nu} A^{\mu}(x)
\end{eqnarray}
which derives from the Lorentz condition $\partial_{\mu} \widetilde{A}^{\mu}(x) = 0$. 
This reduces the number of independent components of $\widetilde{A}^{\mu}(x)$ to three.  
No further gauge transformation can reduce the number of solutions  to below
that, as follows from general field theoretical arguments for massive spin-1 vector gauge bosons, satisfying the above Proca equation for electrodynamics that have lost their gauge invariance because of the mass term.

\section{Derivation of the London equations for a charged condensed Bose liquid 
and of its electro-magnetic properties} 
In the previous two sections we have addressed the emergence of the super-current 
and its associated to it massive electromagnetic field from two diametrically 
opposite points of view:

(i) From the classical phenomenological point of view, based on the London 
equations, in which the steady state of the diamagnetic super-current arises 
from a qualitative modification of the electromagnetic field, which is  
irrevocably locally coupled to this current. Its frictionless transport can 
only occur in conjunction with Meissner-Ochsenfeld screening, which not only results from this current but sustains it.

(ii) From a quantum field theory point of view, in which a symmetry broken 
quantum vacuum constitutes a matter field, whose collective excitations (the 
Nambu-Goldstone modes) are coupled to the quantized excitations of the 
electromagnetic field in which the superconductor matter field is embedded. 
The correlated dynamics of the  charge current of the  matter field and  
the electromagnetic spin-1 vector boson gauge field transforms, via the 
Anderson-Higgs mechanism, the spin-0 scalar Nambu-Goldstone  mode into 
a massive electromagnetic helicity-0 (longitiudinal) spin-1 vector gauge boson.

In the classical description, in terms of the London equations, the concept of a 
phase of the superfluid is totally absent.  In spite of that,  as we have seen in Sec. 
IV, this approach ultimately leads to transform the Maxwell equations into  
Proca equations for massive electromagnetic field excitations which locally 
accompany the super-current and thereby break electromagnetic gauge invariance. 
In the field theoretical description one  assume from the outset a fixed phase 
of the macroscopic Ginzburg-Landau wave function for the superconducting 
ground state before the electromagnetic field is switched on. The necessity of 
such an assumption and thereby of a SSB scenario for a solid state system 
has been questioned since superconductivity  is characterized by an out of 
equilibrium state flow of the super-current, which ultimately relies  on the sole 
fact that the electromagnetic field 
accompanying it has become massive.  Phil Anderson \cite{Anderson-1986} and Anthony 
Leggett together with Fernando Sols, \cite{Leggett-Sols-1991} have addressed
the question to what extent the phase-locking of a superfluid condensate is a 
necessary or even a "meaning full concept" for an isolated superconductor.

A similarly ambiguous situation was encountered when, with Walter 
Thirring \cite{Ranninger-Thirring-1963} in 1962, we derived the London equations 
within a linear response function approach for a paradigm of superconductivity:
the charged Bose liquid. At that memorable time, the forthcoming of fundamentally
new concepts in physics was at its hight. Although it addressed primarily the field theoretical aspects of superconductivity, foremost by Goldstone, Weinberg and Salam\cite{Goldstone-1961,Goldstone-1962}, by Nambu and Jona-Lasinio \cite{Nambu-1960,Nambu-JonaLasinio-1961a,Nambu-JonaLasinio-1961b} and by Schwinger \cite{Schwinger-1962} it was widely shared in the physics community at large, which 
was at ease to navigate between solid state, nuclear and particle physics. Simultaneously, crucial experimental results a that time pinned down the  physical significance of the phase of the superconducting condensate in terms of  
flux quantization \cite{Deaver-Fairbank-1961, Doll-Naebauer-1961} and the 
Josephson tunneling. \cite{Josephson-1962} 

What we found in our study with Walter Thirring was that independent 
on whether these bosons are interacting or not, 
one arrives at the same London equations. In the first case, the linear 
in momentum Bogoliubov modes of the interacting bosons suggest the existence of 
Nambu-Goldstone modes and  hence a SSB  of a continuous phase variable of 
the condensate. In the second case, their free particle like spectrum simply makes
the bosons condense into a  Bose-Einstein condensate for which the phase of 
the particles is not fixed. It merely requires  that the single-particle 
density matrix has a single macroscopic eigenvalue.

Let us now sketch the physics which we unearthed in this work and extend it 
in order to bring it into line with the field theoretical approach to 
superconductivity, discussed in the previous section.

The Hamiltonian for an interacting Bose liquid, exposed to an electromagnetic 
vector field ${\bf A}({\bf x},t)$ is given by $H = H_0 + H_{\bf A}$, with
\begin{eqnarray}
H_0 &=& \int d^3 x \psi^{\dagger}({\bf x},t)
\left(-{\hbar^2 \nabla^2 \over 2M}\right) \psi({\bf x},t) - {1 \over 2}g \int d^3 x d^3 y \psi^{\dagger}({\bf x},t)\psi^{\phantom{\dagger}}({\bf x},t) v({\bf x}- {\bf y}) \psi^{\dagger}({\bf y},t)
\psi^{\phantom{\dagger}}({\bf y},t)
\label{eq:Hamiltonian-0} \\
H_A &=& {e \over 2 M c} \int d^3 x \left[\psi^{\dagger}({\bf x},t) 
\left({\hbar \over i}\right) \nabla \psi({\bf x}, t) - \left({\hbar \over i}\right) \nabla \psi^{\dagger}({\bf x},t) \psi^{\phantom{\dagger}}({\bf x},t)
\psi({\bf  x},t)\right]
{\bf A}({\bf x},t) +  {e^2 \over 2 M c^2} {\bf A}^2({\bf x}, t)\rho({\bf x}, t).  \qquad\qquad\qquad
\label{eq:Hamiltonian-A}
\end{eqnarray}

In  a typical setting  of a Meissner-Ochsenfeld experiment, the circulating 
current in a superconducting ring or a hollow metallic sphere, is triggered by a 
magnetic field. It implies the action of exclusively the 
transverse component of the electromagnetic vector field ${\bf A}({\bf x},t)$, given 
by the magnetic  field ${\bf B}({\bf x},t) = \nabla \times {\bf A}({\bf x},t)$. 
Localized charges, which potentially could act as sources for an electric field
${\bf E}({\bf x},t) = -\frac{1}{c}\partial_t{\bf A}({\bf x},t) - 
\nabla \phi({\bf x},t)$ are absent in such a setup. The  scalar potential 
$\phi({\bf x},t)$ of the electromagnetic field thus can be put equal 
to zero, together with the constraint for the electromagnetic 
gauge: $\partial_t\Lambda({\bf x},t) = 0$. 

The current of the bosonic charge carriers is composed of a magnetic field-induced kinetic
contribution $\langle{\bf J}_{kin}({\bf x},t)\rangle_{\bf A}$ and  a diamagnetic contribution, 
deriving from the associated to it induced charge density fluctuations 
$e \rho({\bf x},t)=\psi^{\dagger}({\bf x},t)\psi^{\phantom{\dagger}}({\bf x},t)$, as given by 
\begin{eqnarray}
&& {\bf J}({\bf x},t) = {\bf J}_{kin}({\bf x},t) -
{e^2 \over M c} {\bf A}({\bf x},t)\rho({\bf x}, t),\\
{\bf J}_{kin}({\bf x},t) &=& -{e \over 2 M} \left[\psi^{\dagger}({\bf x},t) 
\left({\hbar \over i}\right)\nabla \psi({\bf x}, t) - 
\Psi({\bf x},t)\left({\hbar \over i}\right)\nabla \psi^{\dagger}({\bf x},t)\right]. 
\end{eqnarray}
That latter is correlated to the fluctuations of the induced current via the 
continuity equation and controls charge current conservation.

The driving mechanism for achieving a self-sustained conserved "super-current" 
is contained in the linear response of the  Bose condensed system to an
associated to it local source field ${\bf A}({\bf x},t)$
 \begin{eqnarray}
\langle J^m({\bf x},t) \rangle_{\bf A} &=& -{i \over \hbar}
\int ^t_{- \infty} d^3 x' dt'\langle [J^m_{kin}({\bf x},t),H_{\bf A}({\bf x'},t')]\rangle 
- {e^2 \over M c} A^m({\bf x}, t)\langle\rho({\bf x}, t)\rangle \nonumber \\
&=& \int^t_{- \infty} d^3 x' dt'K^{mn}({\bf x}- 
{\bf x'}, t -t')A^n({\bf x}',t') 
\label{eq:current}\\
K^{mn}({\bf x}- {\bf x}', t -t') &=& {i \over \hbar c}\Theta(t -t')
\langle[J^m_{kin}({\bf x},t),J^n_{kin}({\bf x}',t')]\rangle
-{e^2 \over M c}\delta^{mn} \delta({\bf x} -{\bf x}')\delta(t - t')\langle
\rho({\bf x},t) \rangle,
\label{eq:kernel_1}
\end{eqnarray}
similar to Phil Anderson's treatment of the electron plasma, 
\cite{Anderson-1963} discussed in the previous section V. 
In order for the source field ${\bf A}({\bf x}, t)$ to be sustained by the flow 
of the conserved total current ${\bf J}({\bf x},t)$ it has to satisfy the  
Maxwell wave equations with a source term $\langle{\bf J}({\bf x},t)\rangle_{\bf A}$,  satisfying the equation

\begin{equation}
\left[\nabla^2 - {1 \over c^2} \partial_t^2 \right] {\bf A}({\bf x},t) = 
- \langle{\bf J}({\bf x},t)\rangle_{\bf A}.
\label{eq:Maxwellwaveequation}
\end{equation}
Charge conservation in this feed back process between the induced 
kinetic current  $\langle{\bf J}({\bf x},t)\rangle_{\bf A}$ and  the intrinsically diamagnetic
one, $(e^2/Mc){\bf A}({\bf x}, t)$, is assured by relating them via the 
continuity equation. This  was evidently missing  in Felix
Bloch's  initial proposition, Eq. \ref{eq:Londoneq_2ter} in Sec. IV, but was 
introduced in terms of a contribution ${\bf J}_{coll} =\nabla  f({\bf x},t)$ coming from collective
modes in the superconducting state by Nambu, \cite{Nambu-1960} as we have illustrated.

In our treatment of the London equations in ref. \onlinecite{Ranninger-Thirring-1963} we assured a manifestly
gauge invariant description of the charged  Bose fluid, embedded in an externally  applied electromagnetic 
field by imposing the longitudinal sum rule, which derives 
from  the canonical commutation relations 
\begin{eqnarray}
 \langle[J^m_{kin}({\bf x},t), \rho({\bf x}',t'] \rangle &=& - {e \over 2 M}
 \left({\hbar \over i} \right){d \over d x_m} \delta({\bf x} -{\bf x}')\delta(t - t') \langle 2 \rho({\bf x},t)\rangle,\\
 \label{eq:sumrule_1}
  {d \over d x_m}{d\over d x'_n}\langle [J^m_{kin}({\bf x},t), J^n_{kin}({\bf x}',t')] \rangle 
 &=& - {e^2 \over 2 M}
 \left({\hbar \over i} \right)\nabla^2 \delta({\bf x} -{\bf x}')
 \partial_{t'}\delta(t - t') \langle 2 \rho({\bf x},t)\rangle. 
 \label{eq:sumrule_2}
\end{eqnarray} 

In order to render transparent the mutual inter-relation between the conserving current 
of an incipient superconductor (in the absence on any electromagnetic field) and 
the associated to it electromagnetic field, when it interacts with the charge 
carriers, we separate the kernel $K^{mn}({\bf x},t)$ in  Eq. \ref{eq:kernel_1} into 
its longitudinal ($\propto {\bf A}({\bf x},t)$) and transverse 
($\propto \nabla \times \nabla \times {\bf A}({\bf x},t)$) response: 
\begin{eqnarray}
K^{m n}({\bf q}, \varepsilon) &=& \delta^{mn}{e^2 \over 2 M c} \int d \varepsilon'[f_{\parallel}({\bf q},\varepsilon') - f_{\parallel}({\bf q},-\varepsilon')]\left[{1 \over \varepsilon' - \varepsilon - i \eta} - {1 \over \varepsilon' - i \eta} \right] \nonumber \\
&+& (q^m q^n - q^2 \delta^{m n}){e^2 \over 2 M c}
\int d \varepsilon' \left[ f_{\perp}({\bf q},\varepsilon')  
-f_{\perp}({\bf q},-\varepsilon') \right]
\left[{1 \over \varepsilon' - \varepsilon - i \eta}\right].
\label{eq:kernel_2}
\end{eqnarray} 
Expressing the boson wave functions in terms of boson creation and  annihilation operators $\psi({\bf x},t) = \Omega^{-\frac{1}{2}}\sum_{\bf k} b^{\phantom{\dagger}}_{\bf k} e^{-i {\bf k} \cdot {\bf x}},\,\psi^*({\bf x},t) = \Omega^{-\frac{1}{2}}\sum_{\bf k} b^{\dagger}_{\bf k} e^{i {\bf k}\cdot {\bf x}}$ (with $\Omega$ denoting the normalization volume), we have 
\begin{eqnarray}
f_{\parallel}({\bf q},\varepsilon)  &=& {4 \over q^2} \sum_{{\bf k},{\bf k}',z,z'}
{1 \over Z}
e^{-{(E_z -\mu N) \over k_B T}}({\bf q}\cdot{\bf k}){(\bf q}\cdot{\bf k}') \nonumber \\
&& \times\langle z|b^{\dagger}_{{\bf k} - \frac{1}{2}{\bf q}}
b^{\phantom{\dagger}}_{{\bf k} + \frac{1}{2}{\bf q}}|z'\rangle 
\langle z'|b^{\dagger}_{{\bf k}' + \frac{1}{2}{\bf q}}
b^{\phantom{\dagger}}_{{\bf k}' - \frac{1}{2}{\bf q}}|z\rangle \delta (\varepsilon -E_z +E_z') \label{eq:kernel_longitudinal}\\
f_{\perp}({\bf q},\varepsilon)  &=& {2 \over q^2} \sum_{{\bf k},{\bf k}',z,z'}
{1 \over Z}e^{-{(E_z -\mu N) \over k_B T}}
\left({3({\bf q}\cdot{\bf k})({\bf q}\cdot{\bf k}')\over q^2}  - 
({\bf k}\cdot{\bf k}')\right) \nonumber\\
&& \times\langle z| b^{\dagger}_{{\bf k} - \frac{1}{2}{\bf q}}
b^{\phantom{\dagger}}_{{\bf k} + \frac{1}{2}{\bf q}}|z'\rangle
\langle z'|b^{\dagger}_{{\bf k}' + \frac{1}{2}{\bf q}}
b^{\phantom{\dagger}}_{{\bf k}' - \frac{1}{2}{\bf q}}|z\rangle \delta (\varepsilon - E_{z'} +E_z)
\label{eq:kernel_transverse}
\end{eqnarray}
The second term in the brackets for the longitudinal component of the kernel 
in Eq. \ref{eq:kernel_2} derives from the purely diamagnetic contribution to 
the current. Exploiting the longitudinal sumrule, Eq. \ref{eq:sumrule_1}, we 
cast it into  the form
\begin{equation}
2 \langle \rho({\bf x},t)\rangle = {e \over 2M}\int{d \varepsilon' \over \varepsilon'}\left(f_{\parallel}({\bf q},\varepsilon') - 
f_{\parallel}({\bf q},-\varepsilon')\right).
\end{equation}

The first term in Eq. \ref{eq:kernel_2} describes the response of the induced 
current to its generating electric source field ${\bf E}({\bf x},t)$  and the second 
term to its generating magnetic source field ${\bf B}({\bf x},t)$. This 
formulation satisfies the longitudinal and transverse sumrules. \cite{Anderson-1958bis} 
Eq. \ref{eq:kernel_2} is a generalization of an earlier formulation by 
Schafroth, \cite{Schafroth-1955} who focused on the effect of 
Meissner-Ochsenfeld screening in the static limit and therefore restricted himself 
to the London gauge, i.e., $\nabla {\bf A}({\bf x},t) \equiv 0$. Within such a 
choice of electromagnetic gauge fixing, the longitudinal response, which 
introduces the longitudinal massive photons, can not be obtained.
 
The matrix elements in the kernel Eq. \ref{eq:kernel_2} for the Bose liquid, 
described by the Hamiltonian Eqs. \ref{eq:Hamiltonian-0} and 
\ref{eq:Hamiltonian-A} have  been evaluate in our paper, ref. 
\onlinecite{Ranninger-Thirring-1963}, for the free Bose liquid, the weakly 
interacting one and for electrons described by the BCS scenario, after having imposed gauge currents to render the BCS Hamiltonian gauge invariant. 
For the non-interacting free Bose liquid  ($g = 0$ in the Hamiltonian, Eq. \ref{eq:Hamiltonian-0}), 
the space-time Fourier transform of the kernel becomes
\begin{eqnarray}
K^{mn}({\bf q},\varepsilon) =  {e^2 \over 2 M c}\sum_{\bf k} 
{N_{{\bf k} - \frac{1}{2}{\bf q}} - N_{{\bf k} + \frac{1}{2}{\bf q}} \over E_{{\bf k} + 
\frac{1}{2}{\bf q}}- E_{{\bf k} - \frac{1}{2}{\bf q}} -\varepsilon}
\left[\delta^{mn} {2 \varepsilon ({\bf k} \cdot {\bf q}) \over q^2 } + 
\left(\delta^{mn} - {q^m q^n \over q^2}\right)
 2 \left( k^2 - {3 ({\bf k} \cdot {\bf q})^2 \over q^2 }\right)\right]
\label{eq:kernel_3}
\end{eqnarray}
with $E_{{\bf k} \pm \frac{1}{2}{\bf q}} = \frac{\hbar^2 }{2M}({\bf k} \pm \frac{1}{2}{\bf q})^2$
and $N_{{\bf k} \pm \frac{1}{2}{\bf q}} = (exp(E_{{\bf k} \pm \frac{1}{2}{\bf q}} - 
\mu)/k_B T) - 1)^{-1}$ denoting the Bose distribution 
function for a boson density controlled by the chemical potential $\mu$. Given 
the feature of Bose-Einstein condensation with a density of zero momentum 
bosons $N_0$, we substitute $N_{{\bf k} \pm \frac{1}{2}{\bf q}}$ in Eq. \ref{eq:kernel_3}  by $N_0(T)\delta_{{\bf k} \pm \frac{1}{2}{\bf q},0} + 
\widetilde N_{{\bf k} \pm \frac{1}{2}{\bf q}}(T)$. The second term represents the 
non-condensed fraction of the bosons, which for a free non-interacting Bose 
liquid tends to  zero as we approach zero temperature. The expectation value of the current, Eq. \ref{eq:current}, then becomes
\begin{eqnarray}
\langle J^m({\bf q}, \varepsilon)\rangle_A = {N_0 e^2 \over M c}\left[\delta^{mn} 
{\varepsilon^2 \over \omega_q^2 -\varepsilon^2}A^n{\bf q},\varepsilon) - 
{\omega_q^2 \over \omega_q^2 -\varepsilon^2}{1 \over c} \left(\delta^{mn} - {q^m q^n \over q^2}\right) 
A^n({\bf q}, \varepsilon)\right],  
\label{eq:Londoncurrent}
\end{eqnarray}
where $\omega_q = (\hbar^2 q^2/2M)$. With $[rot \, rot \, {\bf A}]({\bf q},\varepsilon) = [rot \, {\bf B}]({\bf q},\varepsilon)$ and 
$i \varepsilon {\bf A}({\bf q},\varepsilon) = c \,{\bf E}({\bf q},\varepsilon)$, we find
\begin{eqnarray}
\langle {\bf J}({\bf q}, \varepsilon)\rangle_A &=& {N_0 e^2 \over 2M} 
{i \varepsilon\over \omega_q^2 -\varepsilon^2}{\bf E}({\bf q},\varepsilon) -
{N_0 e^2 \over 2Mc}{\omega_q^2 \over \omega_q^2 -\varepsilon^2} \left({1 \over q^2}\right)[rot \,{\bf B}]({\bf q},\varepsilon) \nonumber \\
&=& \sigma({\bf q},\varepsilon) \, {\bf E}({\bf q},\varepsilon) + \chi({\bf q},\varepsilon)[rot \, {\bf B}]({\bf q},\varepsilon),
\label{eq:Londoncurrent-bis}
\end{eqnarray}
with
\begin{eqnarray}
\sigma({\bf q},\varepsilon) =  {N_0 e^2 \over 2M}  {i \varepsilon \over \omega_q^2 -\varepsilon^2}, \qquad \chi({\bf q},\varepsilon) = -{N_0 
e^2 \over M c} \left[{1 \over q^2}\right] 
{\omega_q^2 \over \omega_q^2 -\varepsilon^2} 
\end{eqnarray}
denoting the electric conductivity and magnetic susceptibility respectively. Their limiting behavior 

\begin{eqnarray}
\sigma(q = 0, \varepsilon \rightarrow 0 ) = D \delta(\varepsilon), \quad D/ e^2 = {N_0 \over M} \pi \\
\chi(q \rightarrow 0,\varepsilon = 0) = - D_s{1 \over q^2}, \quad D_s/e^2 = {N_0 \over M c}  
\end{eqnarray}
provides us with (i) the Drude weight  $D$, given in terms of the  density to 
mass ratio of itinerant charge carriers and (ii) the superfluid weight $D_s$ in
terms of the  density to  mass ratio of the superfluid charge carriers.  For 
the non-interacting free Bose liquid
we obtain at $T=0$ equal weights $N_0/M$ for $D$ and $D_s$, which is the hallmark
of its superfluid ground state. \cite{Scalapino-Wight-Zhang-1993} As one approaches 
the superfluid transition temperature, $N_0$ tends to zero. $N_{\bf k}$ is then 
given exclusively by the distribution function of the non-condensed bosons 
$\widetilde N_{\bf k} = (exp(E_{\bf k}/k_B T) - 1)^{-1}$. In the long wavelength 
limit, the weight of the transverse component of the kernel, controlling the 
singular behavior of the static ($\varepsilon = 0$) magnetic susceptibility then 
tends to zero in the angle averaged kernel as
\begin{eqnarray}
lim_{q \rightarrow 0}\sum_{\bf k} {N_{{\bf k} - \frac{1}{2}{\bf q}} - N_{{\bf k} + \frac{1}{2}{\bf q}} \over E_{{\bf k} + 
\frac{1}{2}{\bf q}}- E_{{\bf k} - \frac{1}{2}{\bf q}} -\varepsilon}\left({ k^2 q^2 - 3 ({\bf k} \cdot {\bf q})^2 \over q^2 }\right) = 0.
\end{eqnarray} 
The contribution to the kernel arising from the longitudinal part, 
Eq. \ref{eq:kernel_longitudinal}, on the contrary, remains 
finite with a Drude weight, which characterizes a Bose metal state above $T_c$.

The expression for the super-current
$\langle {\bf J}({\bf q}, \varepsilon)\rangle_A$ in Eq. \ref{eq:Londoncurrent} 
contains all the ingredients  necessary for deriving the London equations. 
Multiplying it by $i \varepsilon \equiv \partial / \partial_t$ 
(giving us its time derivative), respectively applying the operator $-i{\bf q}\times \equiv rot$ 
(giving us its rotational component), we obtain
\begin{eqnarray}
i \varepsilon \langle J^m({\bf q},\varepsilon) \rangle_A &=& {N_0 e^2 \over M}
\left( \delta_{mn} - {\omega^2_q \over \omega^2_q - \varepsilon^2}{q_mq_n \over q^2}\right) i \varepsilon 
E^n({\bf q},\varepsilon) = {N_0 e^2 \over M}E^m({\bf q},\varepsilon)
\label{eq:London1-bis} 
\end{eqnarray}

\begin{eqnarray}
-i[{\bf q} \times \langle {\bf J}({\bf q}, \varepsilon)\rangle_A]^m &=& {N_0 e^2 \over M c}
\left(- \delta_{mn} - {\omega^2_q \over \omega^2_q - \varepsilon^2}{q_mq_n \over q^2} \right)
B^n({\bf q}, \varepsilon) = -{N_0 e^2 \over M c}B^m({\bf q},\varepsilon)
\label{eq:London2-bis}
\end{eqnarray}
The second term in the brackets of these equations vanishes because of  
$q_nE^n({\bf q},\varepsilon) = \nabla {\bf E}({\bf q},\varepsilon) = 0$ and  
$q_nB^n({\bf q},\varepsilon) = \nabla {\bf B}({\bf q},\varepsilon) = 0$.
These equations present the generalizations of the classical London equations,
Eqs. \ref{eq:Londoneq_1} and \ref{eq:Londoneq_2}, illustrated in Sec. IV, which 
extend their range of validity beyond that of steady state homogeneous super-
currents to alternating currents with a wavelength $q^{-1}$ and 
frequency $\varepsilon$. It permits us to illustrate the 
emergence of massive photons of the electromagnetic field which accompanies 
these currents. In order to see that let us act onto Eq. \ref{eq:London1-bis} 
once more with $i \varepsilon$ and on Eq. \ref{eq:London2-bis} with ${\bf q}\times$.
We obtain in this way

\begin{eqnarray} 
\left(\varepsilon^2 - c^2 q^2\right)\langle{\bf J}({\bf q}, \varepsilon)\rangle_A = 
{N_0 e^2 \over M c} \left(-i{\bf q} \times {\bf B}({\bf q}, \varepsilon) + {i \varepsilon \over c} {\bf E}({\bf q}, \varepsilon)\right) = 
{N_0 e^2 \over M c}\langle{\bf J}({\bf q}, \varepsilon)\rangle_A,
\label{eq:Procaeq-J}
\end{eqnarray}
where the last equality arises from the Ampere-Maxwell relation
$i{\bf q} \times {\bf B}({\bf q},\varepsilon) -(i \varepsilon/c) {\bf E}({\bf q}, \varepsilon)= 
-\langle{\bf J}({\bf q}, \varepsilon)\rangle_A$.  Eq. \ref{eq:Procaeq-J} describes 
the Proca equations for massive  spin-1 vector boson electromagnetic excitations. 

Projecting out in Eq. \ref{eq:Londoncurrent} the longitudinal  and transverse components of the 
current $\langle {\bf J}({\bf q}, \varepsilon)\rangle_A$, 
\begin{eqnarray}
J^m_{long} &=& \frac{1}{q^2} q_m q_n \langle J^n({\bf q},\varepsilon)\rangle_A = {N_0 e^2 \over Mc} {\varepsilon^2 \over \omega_{\bf q}^2 - \varepsilon^2}A_{\parallel}^m ({\bf q}, \varepsilon) \\
J^m_{trans} &=& \frac{1}{q^2}(q_mq_n - \delta_{mn}q^2) \langle J^n({\bf q},\varepsilon)\rangle_A = -{N_0 e^2 \over Mc} A_{\perp}^m ({\bf q}, \varepsilon)
\end{eqnarray}
describes the response to the longitudinal and transverse components of the vector  field ${\bf A}({\bf q},\varepsilon)$, parallel and orthogonal to the direction of the  propagation of the  electromagnetic waves, given by
\begin{eqnarray}
A_{\parallel}^m ({\bf q}, \varepsilon) &=& {q_m q_n \over q^2}A^n({\bf q}, \varepsilon) ,\quad\qquad {q_m q_n - q^2 \delta_{mn} \over q^2} A_{\parallel}^m({\bf q}, \varepsilon) = 0 \\
A_{\perp}^m({\bf q}, \varepsilon) &=& {q_mq_n - q^2 \delta_{mn} \over q^2} A^n({\bf q}, \varepsilon), \qquad
q_m A_{\perp}^m ({\bf q}, \varepsilon) = 0.
\label{eq:A-perp-parallel}
\end{eqnarray}

We thus derive the wave equations for the electromagnetic excitations  
\begin{eqnarray}
\left(\varepsilon^2 - c^2 q^2 + {N_0e^2 \over Mc}{\varepsilon^2 \over \omega_q^2 - \varepsilon^2}\right)\langle{\bf A}_{\parallel}({\bf q}, \varepsilon)\rangle_A = 0\\
\left(\varepsilon^2 - c^2 q^2 - {N_0e^2 \over Mc}\right)\langle{\bf A}_{\perp}({\bf q}, \varepsilon)\rangle_A = 0,
\end{eqnarray}
which locally accompany the super-current. In the long wavelength limit, 
the longitudinal and the transverse photons  exhibit identical finite masses  
$m_0 = {\hbar \over c}\sqrt{N_0 e^2 \over Mc}$. 

From the results obtained in this section it follows that a free charged Bose 
liquid, exposed to an external magnetic field, exhibits a resistance-less 
super-current and an associated to it Meissner-Ochsenfeld magnetic field screening.  
The London equations, which we derived here for finite frequency and wave
vector dependend magnetic fields remain the same  for a weakly 
interacting charged Bose liquid. We simply have to replace $\omega_q$ by 
a corresponding linear dispersion $\propto g {\bf q}$, which derives from 
Bogoliubov modes (see for details ref. \onlinecite{Ranninger-Thirring-1963}) 
which do invoke a SSB state of a 
phase-polarized ground state: a macroscopic quantum coherent state with an 
arbitrary but fixed phase of the condensate. Given this situation  that 
the superconducting state derived (i)  on the basis of an apparently spontaneously 
symmetry broken SSB state, as expected for an interacting Bose liquid and (ii) 
on the basis of  a single macroscopically occupied state, as expected for a 
non-interacting bose liquid, brings us back to the old  question 
whether the phase of an isolated superconducting condensate has a real physical meaning. 

Phil Anderson \cite{Anderson-1986} considered in this context two superconductors,
prepared independently and whose condensate phases are consequently arbitrary. Will this lead to a Josephson current controlled by a specific or an arbitray phase difference when they are connected by a super-leak? A related scenario 
concerns a gas of atomic hydrogen prepared in two spin polarized states 
with different phases above their condensation temperature 
$T_{\lambda}$. \cite{Siggia-Ruckenstein-1980} The question then is whether 
such a system, upon lowering the temperature to below $T_{\lambda}$, condenses
into two separate condensates with different phases or into a single uniquely 
defined phase, determined by the difference of their two phases.(See also 
the discussion on that in ref. \onlinecite{Leggett-Sols-1991}).
The question of whether an isolated superconductor has a definite 
physically accessible phase or not remains so far unanswered and with 
it the question of whether  SSB of a polarized Many Body ground state
with a physically meaningful phase is a prerequisite for electromagnetic 
gauge symmetry breaking or not. Our example of the charged free
Bose liquid seem to suggest that a macroscopically occupied single-particle 
state is enough for that.

\section{Summary and outlook} 

The truely profound physics behind the  phenomenon of superconductivity became 
apparent only in the years following the publication of the BCS theory, which had explained the totality of thermodynamic and transport features of 
these materials. We have reviewed and highlighted here the revolutionizing new 
concepts which we had learned from that and which were crucial for the developement 
of the whole of physics ever since. We focused here on the major theoretical issues:

(i)  Gauge fields, which accompany  particle matter-waves in their scattering 
and decay processes by safeguarding the basic conservation laws of nature. 
(ii) Spontaneous breaking of continuous symmetries of quantum vacuums, which 
generates conserved currents which describe collective Nambu-Goldstone modes, associated to conserved quantities. 
(iii) The emergence of massive excitations out of such quantum vacuums arises 
in a natural way through the  gauge 
relations linking the two: The massless Nambu-Goldstone modes and the massless
Gauge bosons,  via the Anderson-Higgs mechanism. It ultimately stipulates the 
existence of a yet to be experimentaly verified Higgs boson, presenting the 
source field from which all mass in our universe should originate.

We owe it to the particle field theory community, for having to a large extent unearthed 
these concepts, which they put to great use in consolidating the quantum gauge 
field theoretical approach  in terms of  the standard model. In return, it
had permitted us to regard superconductivity from a broader view, which has 
become invaluable in adressing its hitherto unsuspected manifestations  in  the  
high $T_c$ cuprates and the cold atomic gases.

Transposing the field-theoretical concepts  onto our traditional solid
state physics vision of superconductivity, we arrive at the following picture: 
(A) The gauge bosons of the quantum field theory play the role of the excitations 
of the electromagnetic field in which the superconductor has to be embedded in
order to produce its fundamental features: the persistant resistance-less 
super-currents. (B) Spontaneous symmetry breaking of the highly polarizable Many 
Body ground states of attractively interacting electrons results in an ensemble of
phase-locked Cooper pairs all around the Fermi surface, together with its 
collective Nambu-Goldstone phase fluctuations. \cite{Anderson-1958,
Carlson-Goldman-1973,Kulik-1981} (C) The Anderson-Higgs mechanism, transforming
massless gauge bosons into massive
ones provides in a superconductor the feedback between the circulating charge 
carrying current and the magnetic field which it induces. Its response back onto 
the current stabilizes that later in form  of its steady state persistent 
resistanceless conduction, which is accompanied by a self-induced  
electromagnetic field component whose photonic excitations are massive - as manifest in
Meisser-Ochsenfeld screening. This feature  presents the most profound aspect of 
the superconducting state: its broken electromagnetic gauge symmetry.

Although by the late sixties, all aspects of superconductivity had been understood,
no significant progress had been made  to find a way to increase its 
critical temperature, which evidently always had been the most important 
practical issue of it. However, in 1986 a new class of superconductors had been 
discovered  \cite{Bednorz-Mueller-1986}, which nowadays can provide us  with 
$T_c$'s of around 150 $K$. It was clear from the begining that these 
superconductors were not of BCS type for which $T_c$ is controlled by 
the binding energy of the collective Cooper pair state, i.e., the gap. This BCS scenario 
prevents $T_c$ to become much higher than about 30 K. \cite{Anderson-Matthias-1964} 
$T_c$ in the new superconductors is controlled by the density to mass ratio of the 
superfluid carriers $T_c \propto n_s/m_s$, \cite{Uemura-1989} and that can  reach 
sizeable values, provided that the effective bosonic charge carriers can 
Bose-Einstein condense into a superfluid  state. 

At one time, a promising idea was to consider systems with strong  
electron-phonon coupling which could lead to a superfluid state of bosonic 
bipolarons - Bipolaronic superconductivity. \cite{Alexandrov-Ranninger-1981} 
But unfortunately Bipolarons seem to be condemned to never exist as mobile Bloch 
states and  prefer to condense in isulating states. 
\cite{Chakraverty-Ranninger-Feinberg-1998} According to a 
generally accepted point of view in the chemist's 
community, \cite{Sleight-1976,Sleight-1991,Matthias-1971} high $T_c$'s should 
be possible to be achieved in strong electron-lattice coupled systems, provided they
display intrinsic local dynamical lattice instabilities. This suggestion led me in 
the early eighties, well before the discovery of the cuprate superconductors, to 
propose the phenomenological Boson-Fermion Model (BFM) for such a situation. In 
essence it describes itinerant electrons in a dynamically deformable background 
of molecular clusters, which momentarily trap electrons in form of resonating pairs, 
similar to Feshbach resonance pairing in atomic gases. On a local level, electrons 
then exist simultaneously  in form of itinerant states $c^{\dagger}_{\bf k} = (1/N)\sum_i exp(i {\bf k}\cdot {\bf r}_i)c^{\dagger}_i$, hopping on and off
such molecular clusters  and states in which they are tightly bound together 
in form of localized bipolarons ($BP^{\dagger}_i$). This manifests itself in form 
of a local quantum superposition of these two configurations: of bonding and
anti-bonding two-particle states
$(1/\sqrt2)[ c^{\dagger}_{i \uparrow}c^{i \dagger}_{\downarrow} \pm BP^{\dagger}_i]$,
coexisting with itinerant  single-particle states $c^{\dagger}_i$. 
\cite{Domanski-Ranninger-Robin-1998,Cuoco-Ranninger-2006,Ranninger-Domanski-2010}  
It is this feature which permits one to achieve  bipolarons 
in phase locked coherent states, which in an underlying metallic cristalline 
solid, can condense into a supefluid state, without that they  have to have any prior
to it intrinsic itinerancy. This feature is protected from the aleas 
of the individually locally fluctuating molecular environment (which forms 
them in the first place) by their being locked up in the collective excitations  of 
the Many Body system. \cite{Ranninger-Robin-Eschrig-1995}

The BFM scenario  presents a play-ground for what is generally termed emerging
phenomena \cite{Anderson-1972,Anderson-2011} in Many Body systems and the
associated quantum protection, \cite{Laughlin-Pines-2000,Anderson-2000} 
which assures their dynamical stability. The laws which govern emerging 
phenomena derive from  "macroscopic conservation laws". They are 
dynamically uncorrelated to the laws which control their constituents on a 
local spatial level and which form a highly degenerate quantum vacuum. The 
local consituents, having lost any individuality manifest themselves in the
collective excitations of the macroscopic system, which "protects them against
the vicissitude of the laws governing the individual constituents."

In the BFM, the local physics is controlled by local phase correlations 
between itinerant pairs of electrons passing momentarily through such local 
molecular sites and bound pairs of them, which forms a highly degenerate 
quantum vacuum, composed of such configurations. It characterizes the pseudogap 
state, respectively the insulating phase of the underdoped parent compound, at 
low temperatures, out of which emerges the superconducting  state upon 
doping, respectively reducing the temperature. 
\cite{Cuoco-Ranninger-2006,Stauber-Ranninger-2007} The local phase 
correlations are in competition with the phase correlation controlling the 
macroscopic physics and which try to link together the bosonic components $BP$ of 
spatially separated local molecular bonding states.  \cite{Cuoco-Ranninger-2004} In 
this process those latter loose their  well defined local spectral individuality: 
a characteristic  Boson-Fermion Duality \cite{Ranninger-Domanski-2010}  
in form of a three peak structure in their local single-particle spectral function.
As  extended  single-particle states are formed out of these individual cluster 
states, the corresponding  single-particle spectral features are no longer 
being describeable in terms of simple pole singulatities, 
\cite{Ranninger-Romano-2010} after having been integrated into the dynamics 
of the collective excitations of the macroscopic system.

The physics inherent in the novel high $T_c$ superconductors is clearly not contained in 
the physics we have learned from the BCS superconductors. New concepts like that of 
emergence and quantum protection and most likely more to come will have to be acquired 
to eventually understand the new superconductors on the level we have understood those 
of BCS systems. 

\section{Acknowledgements}

I would like to thank Henri Godfrin for carefully reading this manuscript and for his comments on the issues presented here.

\end{document}